\documentclass[aps,nopacs,showkeys,nofootinbib,preprint]{revtex4}
\usepackage{epsfig,amssymb,bm,graphicx,color,amsmath,rotating}
\usepackage{xcolor}
\usepackage[percent]{overpic}
\usepackage{mwe,xcolor}
\usepackage{transparent}
\usepackage{amsmath}
\usepackage{framed}
\usepackage{slashed}

\newcommand{\dd}{\mathrm{d}}
\begin{document} \normalsize

\title{Holographic bottomonium formation in a cooling strong-interaction medium
at finite baryon density} 

\author{R.~Z\"ollner, B.~K\"ampfer}
\affiliation{Helmholtz-Zentrum  Dresden-Rossendorf, 01314 Dresden, Germany}
\affiliation{Institut f\"ur Theoretische Physik, TU~Dresden, 01062 Dresden, Germany}
 
\begin{abstract}
The shrinking of the bottomonium 
spectral function towards narrow quasi-particle states
in a cooling strong-interaction medium at finite baryon density  
is followed within a holographic bottom-up model.
The 5-dimensional Einstein-dilaton-Maxwell background is adjusted to lattice-QCD results of
sound velocity and susceptibilities. The zero-temperature bottomonium spectral function is adjusted
to experimental $\Upsilon$ ground-state mass and first radial excitations. At baryo-chemical potential $\mu_B = 0$,
these two pillars let emerge the narrow quasi-particle state 
of the $\Upsilon$
ground state at a temperature of about 150~MeV. 
Excited states are consecutively formed at lower temperatures
by about 10 (20)~MeV for the $2S$ ($3S$) vector states. The baryon density, i.e. $\mu_B > 0$, 
pulls that formation pattern to lower temperatures. At $\mu_B = 200$~MeV, we find a shift by
about 15~MeV.
\end{abstract}


\date{\today}

\maketitle

\section{Introduction} \label{introduction}

The observation of sequential bottomonium suppression
\cite{Chatrchyan:2012lxa,Sirunyan:2017lzi,Sirunyan:2018nsz,Acharya:2018mni,Acharya:2020kls}
in relativistic heavy-ion collisions at LHC has sparked a series of dedicated investigations,
e.g.\
\cite{Aronson:2017ymv,Du:2017qkv,Hoelck:2016tqf,Wolschin:2020kwt,Yao:2018sgn,Yao:2020xzw,Yao:2020xwx,Strickland:2019ukl,Brambilla:2020qwo}.
Such
heavy-quark flavor degrees of freedom receive currently some interest 
as valuable probes
of hot and dense strong-interaction matter produced in heavy-ion collisions at
LHC energies. The information encoded, e.g.\ in heavy quarkonia 
($\bar Q Q = c \bar c$ or $b \bar b$)
observables,  supplements
penetrating electromagnetic probes and hard (jet) probes and the rich flow observables,
thus complementing each other in characterizing the dynamics 
of quarks and gluons up to the final hadronic states 
(cf.\ contributions in \cite{Proceedings:2019drx} for the state of the art). 
Heavy quarks emerge essentially in early, hard processes, that is, they witness
the course of a heavy-ion collision -- either as individual entities or subjects  of
dissociating and regenerating bound states. 
Accordingly, the heavy-quark
physics addresses such issues as charm ($c$, $\bar c$) and bottom ($b$, $\bar b$)
dynamics related to transport coefficients 
\cite{Prino:2016cni,Rapp:2018qla,Xu:2018gux,Cao:2018ews,Brambilla:2019tpt,Brambilla:2020qwo,Song:2019cqz} 
in the rapidly evolving and highly
anisotropic ambient quark-gluon medium
\cite{Chattopadhyay:2019jqj,Bazow:2013ifa}
as well as $\bar Q Q$
states as open quantum systems 
\cite{Katz:2015qja,Blaizot:2017ypk,Blaizot:2018oev,Brambilla:2017zei}. 
The wealth of experimental data from LHC,
and also from RHIC, enables a tremendous refinement of our 
understanding of heavy-quark dynamics.
For a recent survey on the quarkonium physics we refer the interested reader to \cite{Rothkopf:2019ipj}.  

The yields of various hadron species, light nuclei and anti-nuclei 
emerging from heavy-ion collisions at LHC energies 
are well described by the thermo-statistical hadronization model 
\cite{Braun-Munzinger:2018hat,Andronic:2017pug} 
over an interval of nine orders of magnitude. 
The final hadrons and nuclear clusters are determined by two parameters: 
the freeze-out temperature $T_{fo} \approx 155$~MeV 
and a freeze-out volume depending on the system size or centrality of the collision. Due to the
near-perfect matter-antimatter symmetry at top LHC energies the 
baryo-chemical potential $\mu_B$ is exceedingly small, $\mu_B/ T_{fo} \ll 1$. 
While the authors of \cite{Reichert:2020yhx} see a delicate interplay of 
elastic and inelastic hadron reactions as governing principle of the
hadro-chemical freeze-out,
it is argued in \cite{Andronic:2017pug} that the freeze-out of color-neutral
objects happens just in the demarcation region of hadron matter to quark-gluon plasma,
i.e. confined vs. deconfined strong-interaction matter. 
In fact, lattice QCD results report a pseudo-critical temperature of 
$T_{pc} = (156 \pm 1.5)$~MeV \cite{Bazavov:2018mes} and $(158.0 \pm 0.6)$~MeV \cite{Borsanyi:2020fev}
-- values agreeing with the disappearance of the chiral condensates and the maximum of some susceptibilities. 
The key is the adjustment of physical quark masses and the use of 2+1 
flavors \cite{Borsanyi:2013bia,Bazavov:2014pvz}, 
in short QCD$_{2+1}$(phys). Details of the 
coincidence of deconfinement and chiral symmetry restoration are matter of debate 
\cite{Suganuma:2017syi}. 
Reference \cite{Bellwied:2018tkc} advocates
flavor-dependent freeze-out temperatures. 
Note that at $T_{pc}$ no phase transition happens, 
rather the thermodynamics is characterized by a cross-over 
accompanied by a pronounced nearby minimum of the sound velocity.
This situation continues to non-zero baryon density as long as the 
baryo-chemical potential $\mu_B$ is small, $\mu_B / T_{pc} \ll 1$. 

Among the tools for describing hadrons as composite strong-interaction
systems is holo\-graphy. Anchored in the famous AdS/CFT correspondence,
holographic bottom-up approaches have facilitated a successful description of
mass spectra, coupling strengths/decay constants etc.\ of various hadron species.
While the direct link to QCD by a holographic QCD-dual or rigorous top-down
formulations are still missing, one has to restrict the accessible observables 
to explore certain frameworks and scenarios. We consider here a framework
which merges for the first time (i) QCD$_{2+1}$(phys) thermodynamics described 
by a dynamical holographic gravity-dilaton-Maxwell
background and (ii) holographic probe quarkonia.
We envisage a scenario which embodies QCD thermodynamics
of QCD$_{2+1}$(phys) and the emergence of hadron states 
at $T_c$  at the same time.
One motivation of our work is the exploration of a holographic model which is in agreement with the above hadron 
phenomenology in heavy-ion collisions at LHC energies. Early holographic studies
\cite{Colangelo:2012jy,Colangelo:2009ra,Colangelo:2008us} to hadrons at finite temperatures faced the problem of meson melting
at temperatures significantly below the deconfinement temperature $T_{pc}$.
Several proposals have been made 
\cite{Zollner:2016cgc,Zollner:2017fkm,Zollner:2017ggh} to find
rescue avenues which accommodate hadrons at and below $T_{pc}$.
Otherwise, a series of holographic models of hadron melting
without reference to realistic QCD thermodynamics, 
 e.g.\
\cite{Braga:2015lck,Braga:2016wkm,Fujita:2009wc,Fujita:2009ca,Grigoryan:2010pj,Braga:2017bml,Braga:2019xwl,Braga:2017oqw,MartinContreras:2021bis} -- mostly with emphasis on quarkonium melting --,
finds quarkonia states well above, at and below $T_{pc}$
in agreement with lattice QCD results
\cite{Bazavov:2014cta,Kim:2018yhk,Ding:2019kva,Larsen:2019zqv}. 
It is therefore tempting to
account for the proper QCD-related background. 

In the
temperature region $T = {\cal O} (T_{pc})$, the impact of charm and bottom
degrees of freedom on the quark-gluon--hadron thermodynamics is minor \cite{Borsanyi:2016ksw}. 
Thus, we consider quarkonia, in particular bottomonium, as test particles. 
We follow 
\cite{Gubser:2008ny,Finazzo:2014cna,Finazzo:2013efa,Zollner:2018uep} 
and model  the holographic background 
by a gravity-dilaton set-up supplemented by a Maxwell field
\cite{DeWolfe:2010he,DeWolfe:2011ts},
i.e.\ without adding further fundamental degrees
of freedom to the dilaton. 
That is, the dilaton potential and its coupling to the Maxwell field 
are adjusted to QCD$_{2+1}$(phys) lattice data. 
Our emphasis is here on the formation of bottomonium in a cooling
strong-interaction environment. Thereby, the bottomonium properties
are described by a spectral function. 
The primary aim of the present paper is to study the impact of a finite
baryon density of the strong-interaction medium, thus complementing
\cite{Zollner:2020cxb,Zollner:2020nnt}.
Finite baryon effects become relevant at smaller beam energies, 
e.g.\ at RHIC, and are systematically accessible in the beam energy scans 
\cite{Odyniec:2019kfh,Abdallah:2021fzj,Bzdak:2019pkr}.
We restrict ourselves to equilibrium and leave non-equilibrium effects,
e.g.\ \cite{Bellantuono:2017msk,Yao:2017fuc}, for future work.

Such $\mu_B > 0$ effects on holographic bottomonium spectroscopy 
have been considered, e.g.\ in
\cite{Braga:2017oqw,Braga:2019xwl,Braga:2020myi}.
Our present investigation is distinguished by choosing a holographic bottom-up
background which is adjusted to QCD-lattice data of sound velocity and 
susceptibilities in the temperature range 100~MeV $ < T < $ 600~MeV.
That is, the gravity-dilaton-Maxwell fields are dynamically determined 
by solutions of the Einstein equations consistent with the equations of motion
of dilaton and Maxwell fields. We do not touch the large-$\mu_B$ region or
a conjectured critical point \cite{DeWolfe:2010he,DeWolfe:2011ts,Rougemont:2015wca,Critelli:2017oub,Knaute:2017opk,Knaute:2017lll,Grefa:2021qvt}
since the experimental access to bottomonium 
physics is expected to be feasible at not too small beam energies,
i.e.\ at low-values of $\mu_B$ in the central rapidity region.      

Our paper is organized as follows. 
In Section \ref{quarkonium}, the dynamics of the probe 
quarkonia is formulated, and the coupling to the thermodynamics-related background
is explained in Section \ref{EDM}. Both ingredients are joint
in Section \ref{SF} for the 
calculation of the spectral functions. The numerical results for the bottomonium 
states $\Upsilon (1S, 2S, 3S)$ are presented in Section \ref{num_results}.
We summarize in Section \ref{summary}. 
Appendix \ref{z_coordinate} details the field equations for the 
Einstein-dilaton-Maxwell model with radial bulk coordinate $z$.
Appendix \ref{UV_IR} considers some options for UV-IR matching to generate
within holography the $\Upsilon (n S)$ $I^G (J^{PC}) = 0^- (1^{--})$ spectrum.

\section{Bottom-up model for quarkonia}\label{quarkonium}

In thermal equilibrium, the admixture of equilibrated heavy quarks 
in strong-interaction matter at $T < 250$~MeV is small 
\cite{Borsanyi:2016ksw,Bellwied:2015lba}.
Rather, initial hard parton interactions (essentially gluon fusion) create
heavy quarks in heavy-ion collisions. 
Thus, heavy quark pairs serve as test particles and need not to be 
back-reacted. In particular, quarkonia constituents are decoupled 
from the ambient quark content, with the exception of the gluon component.
In a model with minimalistic field content one would prefer to keep the effective
gravity-dilaton background 
(extended by the Maxwell field ${\cal B}$ for mimicking $\mu_B > 0$)
to catch QCD thermodynamics 
and attribute to the test particles solely one vector field ${\cal A}$.
A $U(1)$ gauge field ${\cal A} (z)$ in the bulk is supposed to be 
the dual of the vector meson
current operator $\bar Q \gamma^\mu Q$ at the boundary. 
The string-frame action is accordingly 
\begin{equation} \label{test_action}
S_m^V = \frac{1}{k_V} \int \dd^4x \, \dd z \sqrt{g_5} \, \frac14 e^{-\phi_m} \,F_{\cal A}^2, \quad
\phi_m := \phi - \log G_m (\phi),
\end{equation}
where $F_{\cal A}$ stands for the Abelian field-strength tensor of ${\cal A}$
and $k_V=\frac{N_c}{24\pi^2}$ with number $N_c = 3$ of colors.

In contrast to common previous practice,
the background quantities $g_5$ (metric determinant)
as well as $\phi$ (dilaton field) and ${\cal B}$ (Maxwell field) are universal
for any test particle, therefore, $G_m$ encodes solely the essential properties
of the respective test particle. We attribute the quarkonia masses to the
considered test particle. Rather than including the heavy-quark masses
explicitly, we encode them in the following manner in $G_m$.  From the
the ansatz
${\cal A}_\mu = \epsilon_\mu \, \varphi (z) \, \exp\{ i p_\nu x^\nu \}$
with $\mu, \nu = 0, \cdots, 3$, which uniformly separates the $z$ dependence
of the gauge field by the bulk-to-boundary propagator $\varphi$ for
all components of ${\cal A}$,
and the constant polarization vector $\epsilon_\mu$
and gauges ${\cal A}_z  = 0$ and 
$\partial^\mu {\cal A}_\mu = 0$,
the equation of motion follows from the action (\ref{test_action}) as
\begin{equation} \label{eq:EoM}
\varphi'' +
\left[\frac12 A' + (\partial_\phi \log G_m -1 ) \phi' + (\log f)' \right] \varphi' +
\frac{p^\mu p_\mu}{f^2} \varphi = 0,
\end{equation}
which is cast in the form
of a one-dimensional Schr\"odinger equation with  the tortoise coordinate $\xi$
\begin{equation} \label{eq:6}
\left[\partial_\xi^2 - (U(z(\xi)) - m_n^2) \right] \psi_n (\xi)= 0, 
\quad n = 0, 1, 2,  \cdots ,
\end{equation} 
by the transformation 
$\psi (\xi) = \varphi(z (\xi)) \, \exp\{ \frac 12 \int_0^\xi \dd z  \, {\cal S} (\xi) \}$
and $p^\mu p_\mu \to m_n^2$.
One has to employ $z(\xi)$ from solving $\partial _\xi = (1/f) \partial_z$.
The Schr\"odinger-equivalent potential in (\ref{eq:6}) is  
\begin{equation} \label{eq:7}
U := \left( \frac12 {\cal S}' + \frac14 {\cal S}^2 \right) f^2 
+ \frac12 {\cal S} f f' 
\end{equation}
as a function of $\xi(z)$ with
\begin{equation} \label{eq:G}
{\cal S} :=  \frac12 A' - \phi' + \partial_z \log G_m (\phi(z)).
\end{equation}
A prime means the derivative w.r.t.\ the bulk coordinate $z$.

At $T = 0$, we have $f = 1$ and $\xi = z$, and $m_n$ in Eq.~(\ref{eq:6})
is the quarkonium mass spectrum to be used as input. Therefore, the
Schr\"odinger-equivalent potential $U(z)$ must be chosen in such a manner
to deliver the wanted values of $m_n$.
With given $U(z)$, the Ricatti equation (\ref{eq:7}) must be solved for 
${\cal S}$, which in turn determines the heavy-quark mass-specific function
$G_m (\phi)$ via Eq.~(\ref{eq:G}). This $G_m (\phi)$ is assumed as 
independent of temperature and baryo-chemical potential, 
i.e.\ is ready for direct use at $T > 0$ and $\mu_B > 0$ as well. 

In the described chain of operations for getting $G_m$, 
the zero-temperature background quantities $A(z)$ and $\phi(z)$ are needed. 
They are determined by the temperature independent dilation potential $V(\phi)$,
which is adjusted to lattice-QCD thermodynamics data, briefly recalled in the next section.

\section{Background generated by the Einstein-dilaton-Maxwell bottom-up model} 
\label{EDM}

We follow here closely the Einstein-dilaton-Maxwell (EdM) model 
of \cite{Knaute:2017opk},
see also \cite{Knaute:2017lll,Critelli:2017oub,Grefa:2021qvt}.
The EdM action reads
\begin{equation} \label{EdM_action}
S_{EdM} = \frac{1}{2 \kappa_5^2} \int \dd^4 x \, \dd z \, \sqrt{-g_5}
\left( R - \frac12 \partial^M \phi \, \partial_M \phi - V(\phi) - 
\frac14 {\cal F}(\phi) F_{{\cal B}}^2 \right) + S_{GH},
\end{equation}
where $R$ is the Einstein-Hilbert part,
$F_{\cal B}^{MN} = \partial^M {\cal B}^N - \partial^N {\cal B}^M$
stands for the field strength tensor of Abelian gauge field ${\cal B}$ \`{a} la Maxwell
with ${\cal B}_M \dd x^M = \Phi (z) \, dt$  defining the electro-static potential,
and $\phi$ is a real scalar (dilatonic) field with self-interaction 
described by the potential $V(\phi)$.  
The bulk Maxwell field is sourced by the conserved light-quark baryon-current
$\bar q \gamma^\mu q$ at the boundary. In such manner, this field is related
to baryon density effects, parameterized by $\mu_B$.
The Maxwell field and dilaton are coupled by the dynamical strength function
${\cal F}(\phi)$ \cite{DeWolfe:2010he,DeWolfe:2011ts}.
(Note the analogy of ${\cal F}(\phi)$ in (\ref{EdM_action}) and $e^{- \phi_m}$ in (\ref{test_action}).)
The Gibbons-Hawking term $S_{GH}$ for a consistent formulation 
of the variational problem is not needed explicitly in our context.  
The numerical value of the ``Einstein constant” $\kappa_5^2 = 8 \pi G_N$  
is irrelevant in our context.
The metric determinant $g_5$ is related to the ansatz of the infinitesimal line element
\begin{equation} \label{eq:3}
\dd s^2 = g_{MN} \dd x^M \dd x^N :=
 \exp\{ A(z, z_H)\} 
\left[f(z, z_H) \, \dd t^2 - \dd \vec x^2 - \frac{\dd z^2}{f(z, z_H)} \right] ,
\end{equation}
with warp function $A$ and blackening function $f$, already used in Section \ref{quarkonium}.

We relegate the field equations following from the action (\ref{EdM_action})
in the coordinates (\ref{eq:3})
to Appendix \ref{z_coordinate}, 
but mention here the employed parameterizations
\begin{eqnarray} \label{dilaton_potential}
L^2 V(\phi) &=& \left\{
\begin{array}{l}
-12 \exp\{ \frac12 a_1 \phi^2 + \frac14 a_2 \phi^4 \} \mbox{ for } \phi < \phi_x , \\
a_{10} \cosh [a_4(\phi-a_5)]^{a_3/a_4} \exp\{a_6 \phi+\frac{a_7}{a_8} \tanh [a_8(\phi-a_9)]\}
\mbox{ for } \phi > \phi_x , 
\end{array} \right. \label{dil_pot}\\
{\cal F} (\phi) &=& c_0+c_1 \tanh [c_2(\phi - c_3)] +c_ 4 \exp \{-c_5 \phi \}
\label{dyn_coup}
\end{eqnarray}
and refer the interested reader to \cite{Knaute:2017opk} for listings of the parameters
$a_{1, \cdots, 10}$, $\phi_x$, $c_{0, \cdots, 5}$ etc.
Figures 1 and 2 in \cite{Knaute:2017opk} exhibit the excellent agreement with lattice-QCD data
in the interval $T \in [100, 500]$~MeV and remaining uncertainties due to limited precision,
in particular of the sound velocity in the interval $T \in [100, 160]$~MeV and the susceptibility $\chi_4$. 
The scale setting is accomplished by
$L^{-1} = 5.148$~GeV.\footnote{References  \cite{Zollner:2020nnt,Zollner:2020cxb}
use the dilaton potential $L^2 V(\phi) = - 12 \cosh(\gamma \phi) + \phi_2 \phi^2 + \phi_4 \phi^4$
which delivers for $(\gamma, \phi_2, \phi_4) = (0.568, - 1.92, - 0.04)$,
i.e.\ $\Delta^V = 3.9$
also a good description of lattice-QCD data of $v_s^2 (T)$ by $L^{-1} = 1.99$~GeV.
The difference of these scale settings can be traced back to
the sensibility of internal model quantities, such as the dilaton profile $\phi(z, z_H)$,
while observables remain stable, since effects of different model
parameterizations cancel out when considering observables.
Note that the parameterization (\ref{dil_pot}) implies the conformal dimension $\Delta^V = 3.76$.}
The locus of the minimum sound velocity squared is described in leading order by
$T_{min \{ v_s^2\} } (\mu_B) = T_{v_s^2} (\mu_B = 0) \left[1 - \kappa
\left( \displaystyle \frac{\mu_B}{T_{v_s^2}(\mu_B = 0)} \right)^2 \right]$ with
$T_{v_s^2}(\mu_B = 0) = 145$~MeV and $\kappa = 0.0178$~MeV. 
Note that $T_{v_s^2} (\mu_B =0) < T_{pc}$.
Despite the direct relation to an observabe,
the location of $min \{ v_s^2 \}$ is not so precisely constrained by lattice-QCD data
as that of the maximum of chiral susceptibility
which determines quite accurately $T_{pc}$ 
\cite{Bazavov:2018mes,Borsanyi:2020fev}. In so far, the curves
$T_{min \{ v_s^2\} } (\mu_B)$ and $T_{pc}(\mu_B)$ need not to coincide.

The EdM model with these input data is then ready to transport the 
thermodynamic information 
from $\mu_B = 0$ to $\mu_B > 0$, 
thus uncovering the $T$-$\mu_B$ plane.  
This is very much the spirit of the quasi-particle model 
\cite{Peshier:1999ww,Peshier:2002ww}, 
where a flow equation facilitates such a transport.

\section{Spectral functions}\label{SF}

The equation of motion (\ref{eq:EoM}) of $\varphi$
can also be employed to compute quarkonia spectral functions, 
cf.~\cite{Colangelo:2012jy,Fujita:2009wc,Fujita:2009ca,Grigoryan:2010pj,Hohler:2013vca,Teaney:2006nc}.
For $\omega^2  = p^{\mu} p_{\mu} > 0$ fixed, the asymptotic
boundary behavior facilitates two linearly
independent solutions by considering the leading-order terms
on both sides of the interval $[0,z_H]$.
(i) For $z \to 0^+$, one has 
the general solution 
$\varphi(z \to 0) \to A(\omega) \varphi_1 + B(\omega) \varphi_2$,
due to the AdS asymptotic at the boundary,
with two $\omega$-dependent complex constants $A$ and $B$, 
and $\varphi_1(z \to 0) \to 1$ and  $\varphi_2(z \to 0) \to (z/z_H)^2$.
(ii) Near the horizon, $z \to z_H^-$, the asymptotic behavior of solutions of
(\ref{eq:EoM}) is
steered by the poles of $1/f$ and $1/f^2$. 
The two linearly independent solutions are
$\varphi_{\pm} (z \to z_H) \to (1 - \frac{z}{z_H})^{\pm i \omega /|f'(z_H)|}$,
where $\varphi_{\pm}$ represent out-going and in-falling solutions,
respectively.
The general near-horizon solution is given by
$\varphi(z \to z_H) \to C(\omega) \varphi_(z)+ + D(\omega) \varphi_-(z) $,
again with complex constants $C$ and $D$ which depend on $\omega$. 
The side conditions for 
the bulk-to-boundary propagator are $\varphi(0)=1$, 
which means $A(\omega)=1$, and 
$\varphi(z \to z_H) = \varphi_- (z \to z_H)$
(purely in-falling solution at the black hole horizon), 
yielding $C(\omega)=0$. 
Due to the bilinear mapping $(A,B) \mapsto (C,D)$,  
the value of $B$ for getting the desired
in-falling solution can be determined by solving the above equations
twice, once with $A=1$, $B=0$ and once with $A=0$, $B=1$ 
and comparing the result with $\varphi_-$ to dig out the proper
coefficients. 

The corresponding retarded Green function $\mathcal{G}^{\rm R}$ of
the dual current operator $\bar Q \gamma_\mu Q$,
defined within the framework of the holographic dictionary 
via a generating functional by
$\mathcal{G}^{\rm R} = \frac{\delta^2}{\delta {\cal A}^{0 \, \mu} (-\omega)
\delta {\cal A}^0_\mu (\omega)}
\langle \exp\{i \int \dd^4 x \, {\cal A}^0_\nu \bar Q \gamma^\nu Q \} \rangle $,
is given by
\begin{equation} \label{Green_function}
\mathcal{G}^{\rm R} (\omega) =
\frac{\delta^2 S_m^{V,\, \mbox{on-shell}}}{\delta \mathcal{A}^{0 \, \mu} (-
\omega) \, \delta \mathcal{A}^0_\mu (\omega)} 
= \frac{1}{k_V} \lim \limits_{z \to 0} \frac1z
\varphi^* (z) \varphi' (z)
= \frac{2 }{k_V z_H^2} B(\omega)
\end{equation}
with $\mathcal{A}^0_\mu \equiv \epsilon_{\mu} \exp\{ip_{\nu}x^{\nu}\}$
for $\mu \in \{1, 2, 3 \}$  \cite{Teaney:2006nc}.
The quantity $S_m^{V,\, \mbox{on-shell}}$ denotes here the
action (\ref{test_action}) with the solution $\varphi$ from (\ref{eq:EoM}).
Finally, the spectral function $\rho$ follows from
$\rho (\omega) =  
\mbox{Im} \,  \mathcal{G}^{\rm R} (\omega) = 
\frac{2}{k_V z_H^2} \mbox{Im} \, B (\omega)$.
It has the dimension of energy squared,
suggesting to use $L^2 \rho$ or $\rho / \omega^2$ as convenient representations.

\section{Numerical results}\label{num_results}

The spectral function $\rho (\omega, T, \mu_B)$ is accessible by numerical means
by the following chain of operations: 
(i) solving the equations of motion (\ref{A1} - \ref{A4}) following from the action (\ref{EdM_action})
with boundary conditions (\ref{A5} - \ref{A9})
for the background encoded in $A_0(z)$, $f_0(z) = 1$, $\phi_0(z)$ 
with the prescribed $V(\phi_0)$ from Eq.~(\ref{dilaton_potential})
yields the input for
Eqs.~(\ref{eq:7}, \ref{eq:G}) for the determination of $G_m(\phi)$ at $T = 0$ 
(highlighted by the subscript ``0", using $U_0$ from Appendix \ref{UV_IR}; for parameter values,
see Appendix \ref{Ups_approx}),  
(ii) using afterwards that $G_m(\phi)$  in Eqs.~(\ref{eq:EoM}, \ref{Green_function})
but with $A(z; z_H)$, $f(z; z_H)$, $\phi(z, z_H)$ determined again 
by $V (\phi)$ via the equations of motion(\ref{A1} - \ref{A4})  following from the action 
(\ref{EdM_action}) with boundary conditions (\ref{A5} - \ref{A9}) and $ {\cal F} (\phi)$
from Eq.~(\ref{dyn_coup}), see Appendix \ref{z_coordinate}.
Some care is needed in that numerical treatment.

\begin{figure}[tb!]
\includegraphics[width=0.31\columnwidth]{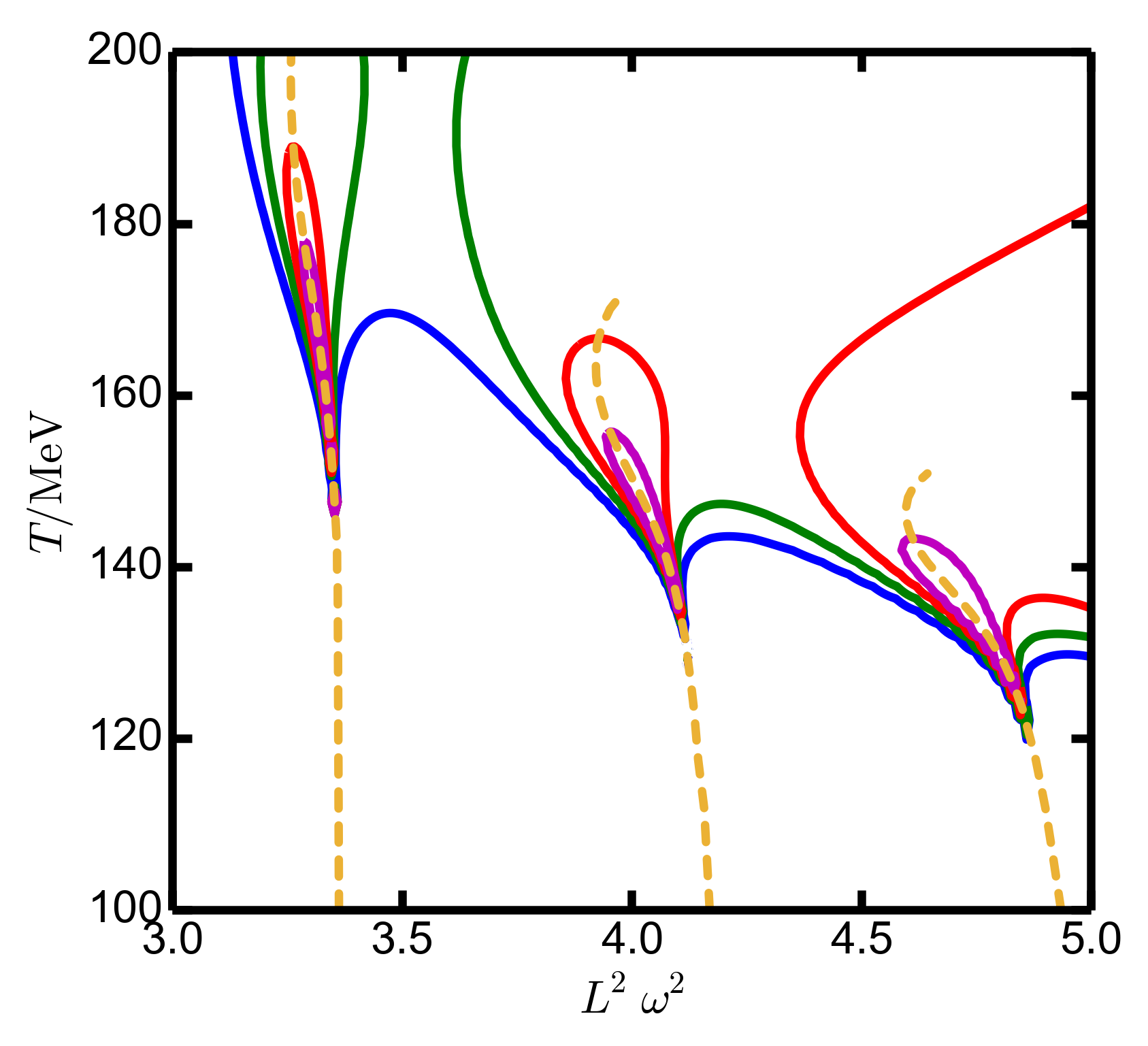} 
\includegraphics[width=0.31\columnwidth]{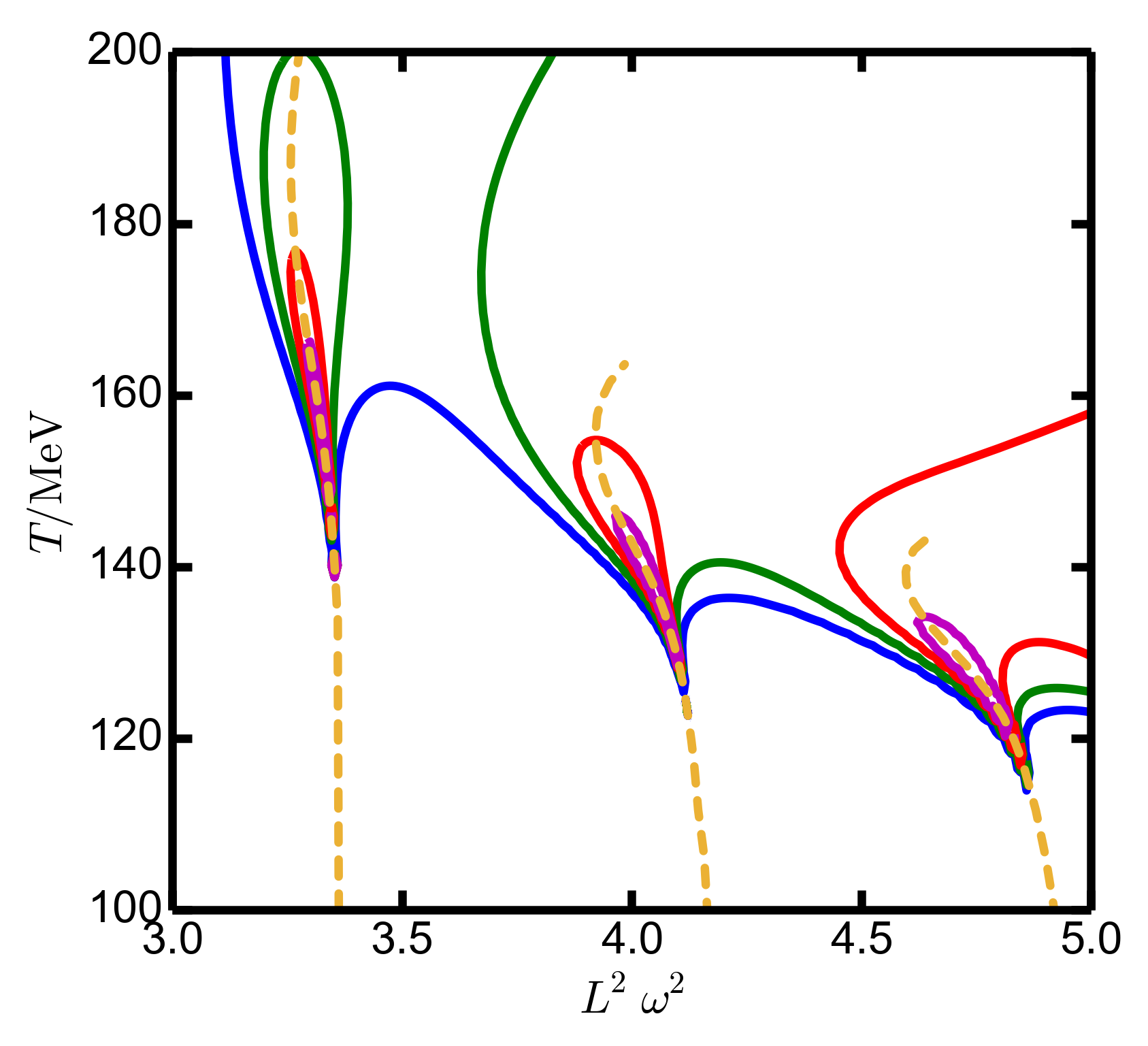} 
\includegraphics[width=0.31\columnwidth]{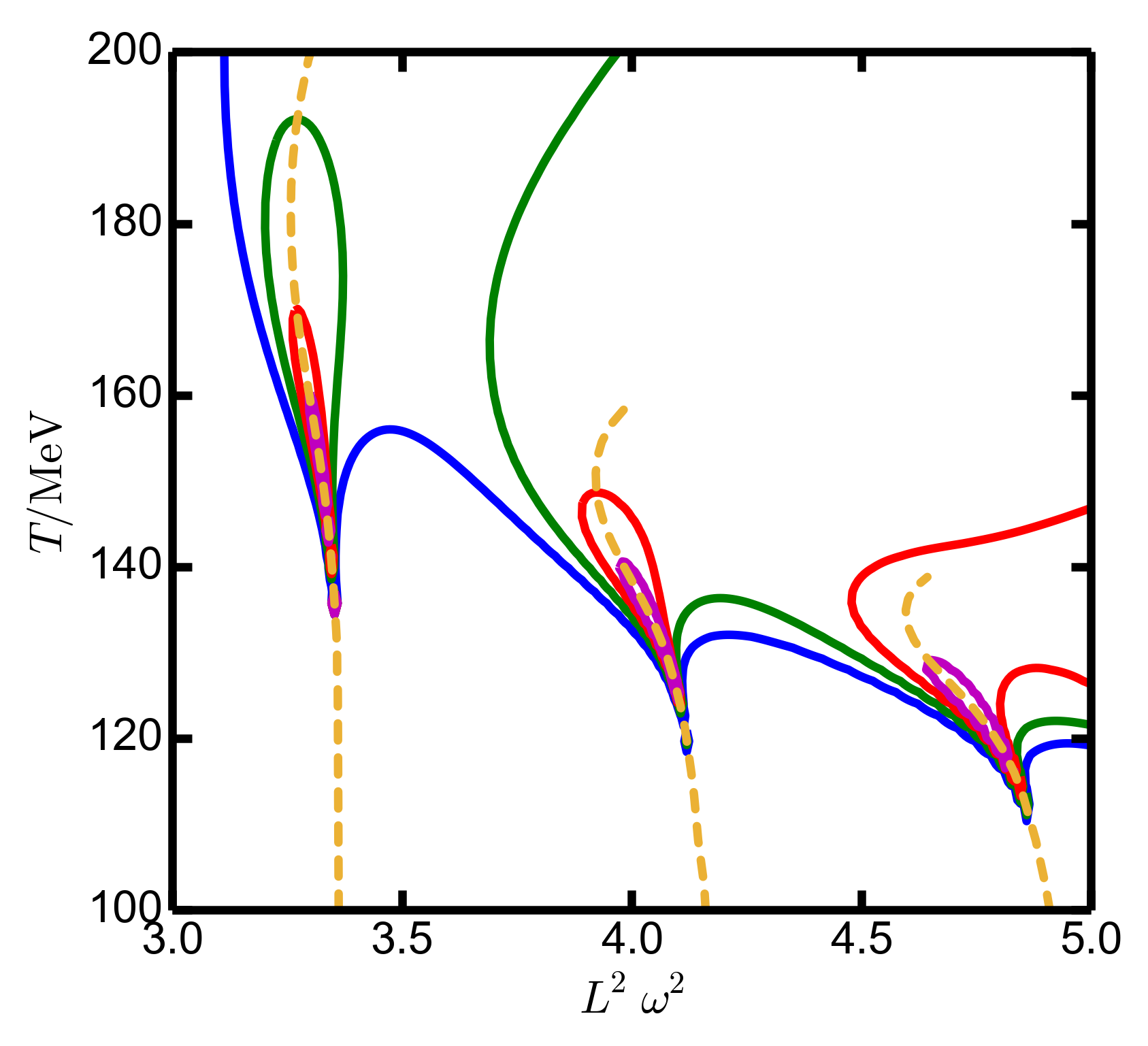} 
\put(-362,107){$\mu_B = 0$}
\put(-247,107){$\mu_B \quad = 100$~MeV}
\put(-97,107){$\mu_B \quad = 200$~MeV}
\caption{Contour plots of the $\Upsilon$ spectral function 
$L^2 \rho (\omega, T, \mu_B)$
over the temperature 
vs.\ scaled frequency 
$L^2 \omega^2$ plane for $\mu_B = 0$ (left), 100~MeV (middle)
and 200~MeV (right);
blue, green, red and magenta contour curves correspond to 
$L^2 \rho = 1$, 3, 10 and 30.
The dashed orange curves mark the maximum positions
of $L^2 \rho (\omega, T, \mu_B)$ w.r.t.\ $\omega$.
At large temperatures, such local maxima
disappear and the respective state can be considered as completely molten. At lower temperatures,
sharp quasi-particles have been formed, i.e.\ the contours are squeezed 
and are hidden behind the dashed orange curves.
\label{SF_Ups} 
}
\end{figure}

\begin{figure}[tb!]
\includegraphics[width=0.31\columnwidth]{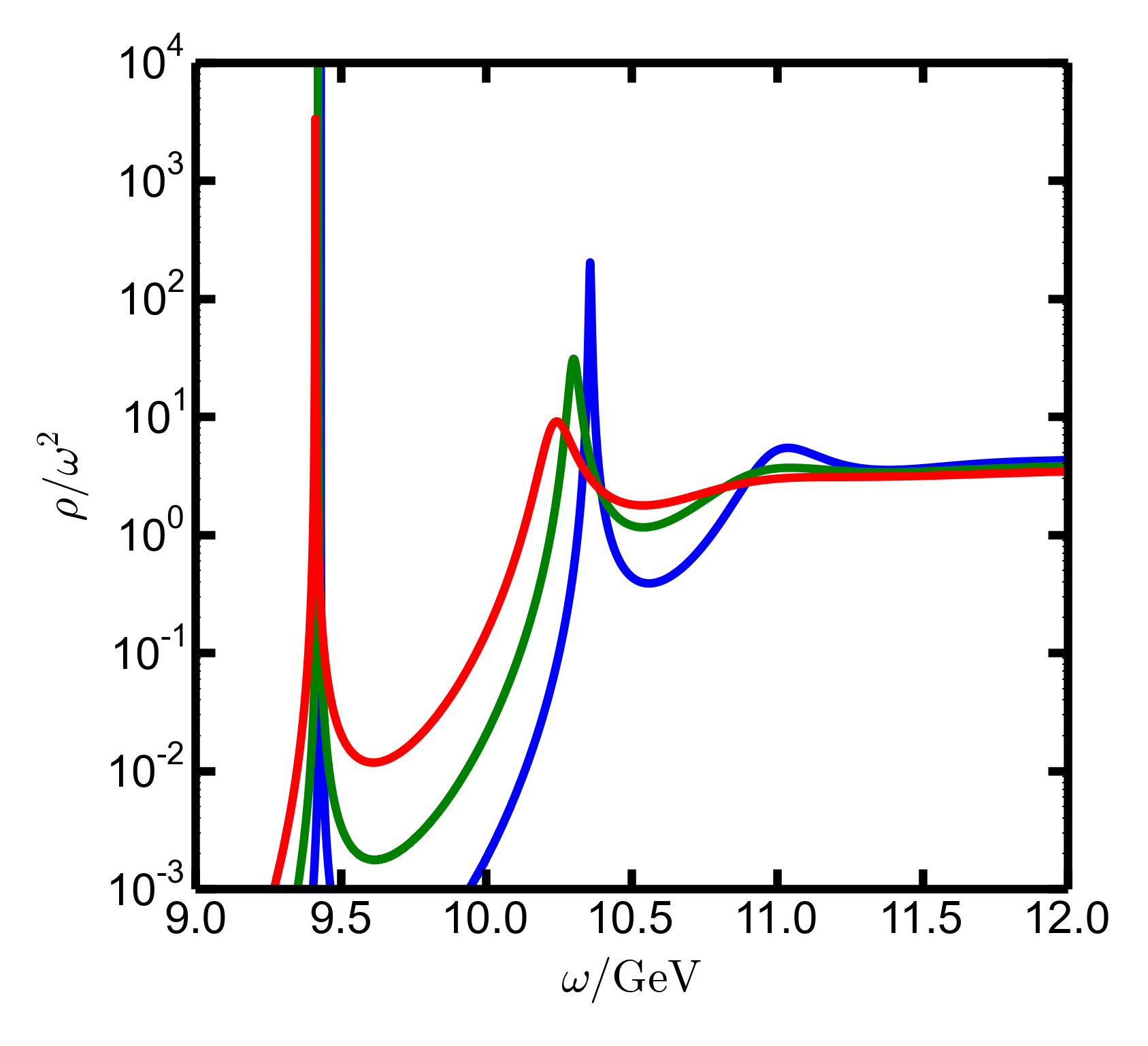}
\includegraphics[width=0.31\columnwidth]{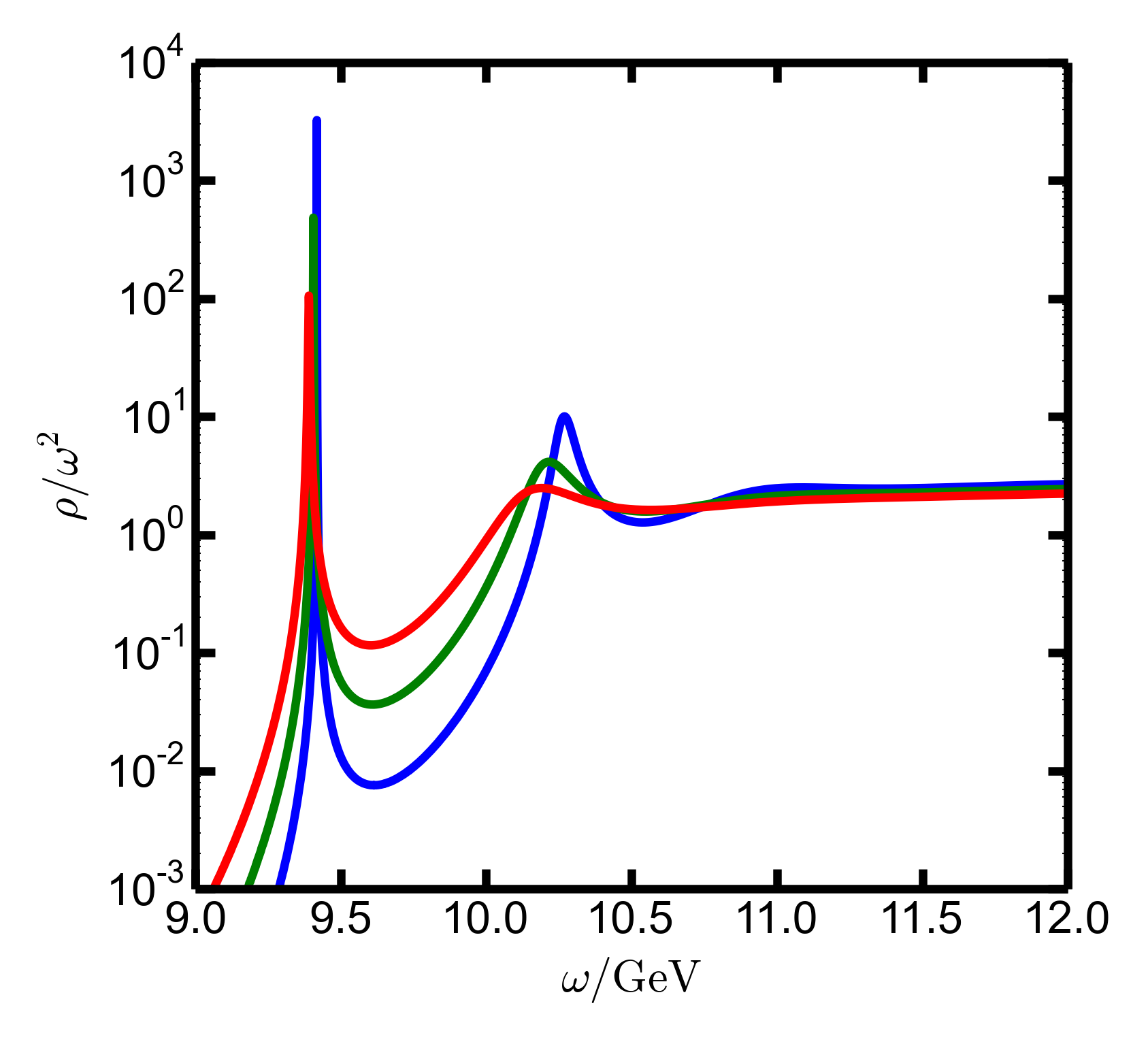}
\includegraphics[width=0.31\columnwidth]{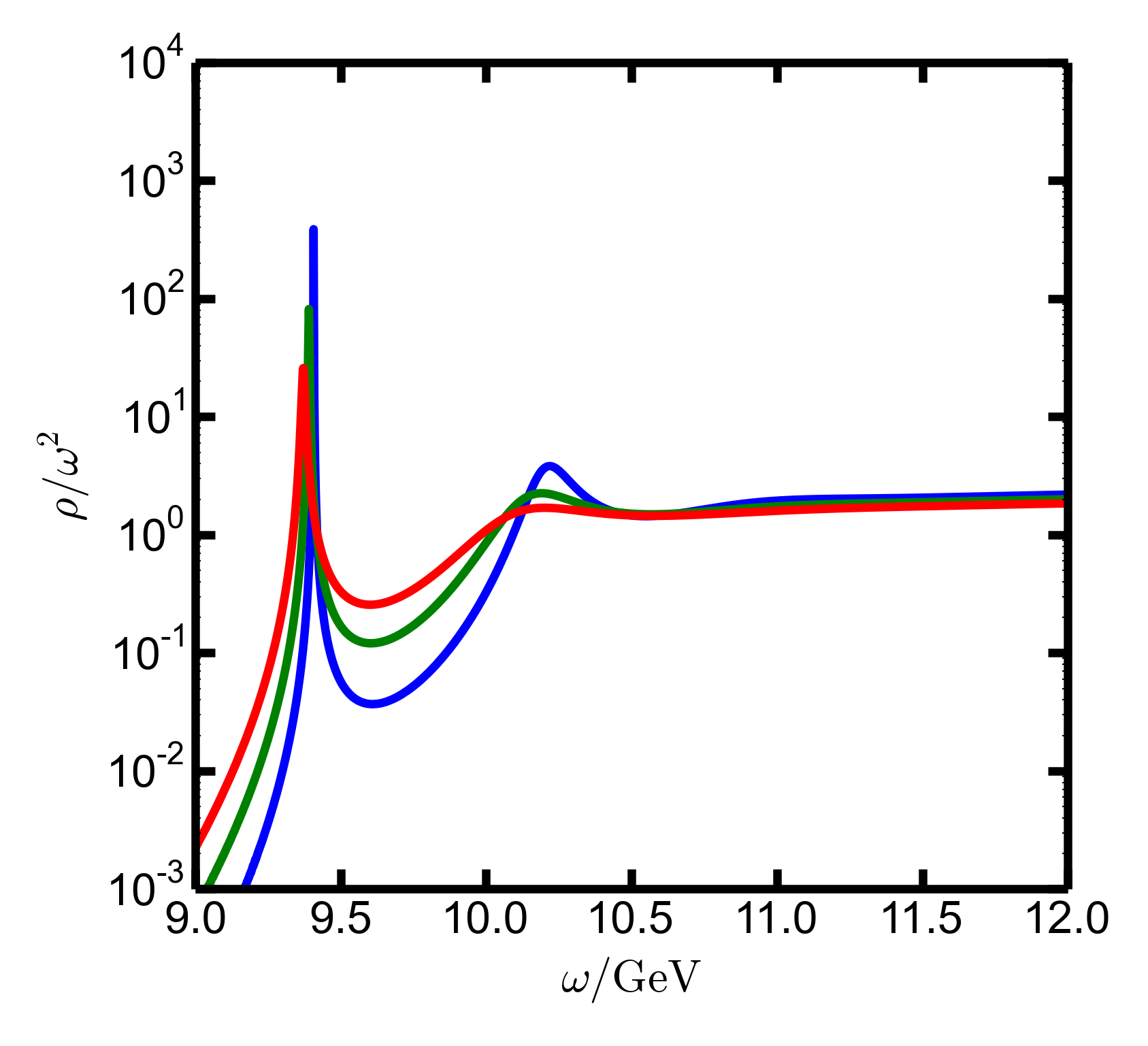}
\put(-350,107){$\mu_B = 0$} 
\put(-240,107){$\mu_B = 100$~MeV} 
\put(-90,107){$\mu_B = 200$~MeV}
\put(-79,50){{\color{red} ----- 155 MeV}}
\put(-79,38){{\color{green} ----- 150 MeV}}
\put(-79,26){{\color{blue} ----- 145 MeV}}
\caption{$\Upsilon$ spectral function $L^2 \rho (\omega, T, \mu_B) $
as a function of scaled frequency 
$L^2 \omega^2$ for $\mu_B = 0$ (left), 100~MeV (middle)
and 200~MeV (right) at 155~MeV (red), 150~MeV (green) and 145~MeV (blue).
These plots arise from Fig.~\ref{SF_Ups} as cross sections at constant
temperature.
\label{SF_Ups_crosssection_T} 
}
\end{figure}

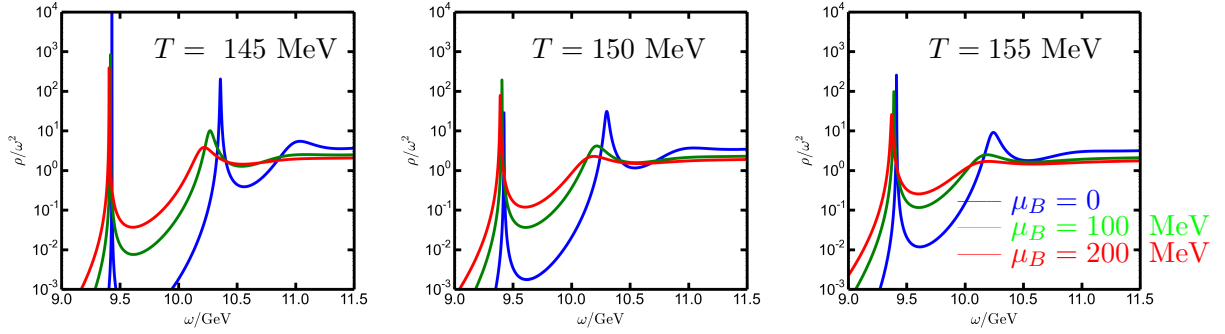
\begin{figure}[tb!]
\includegraphics[width=0.31\columnwidth]{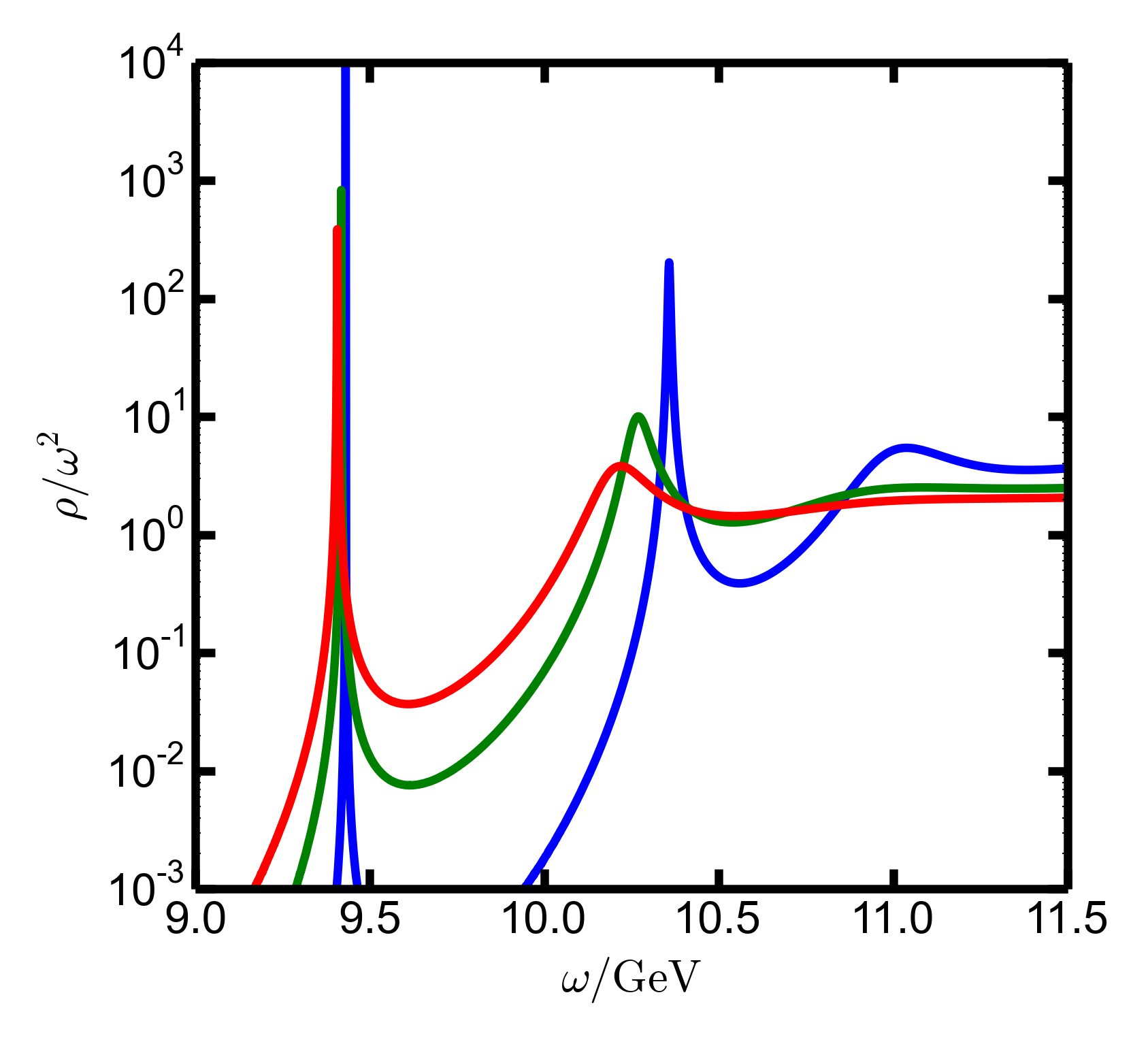}
\includegraphics[width=0.31\columnwidth]{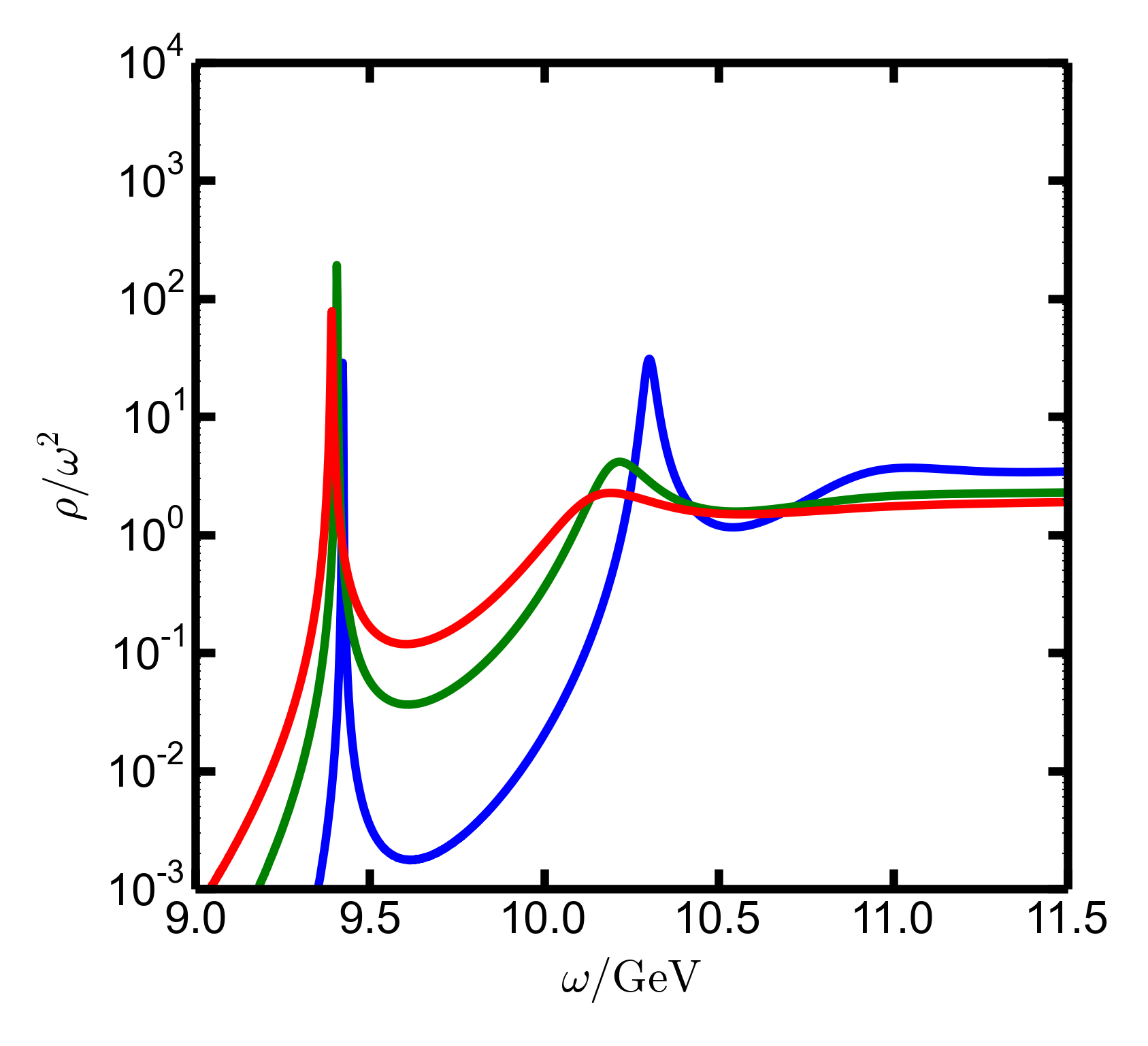}
\includegraphics[width=0.31\columnwidth]{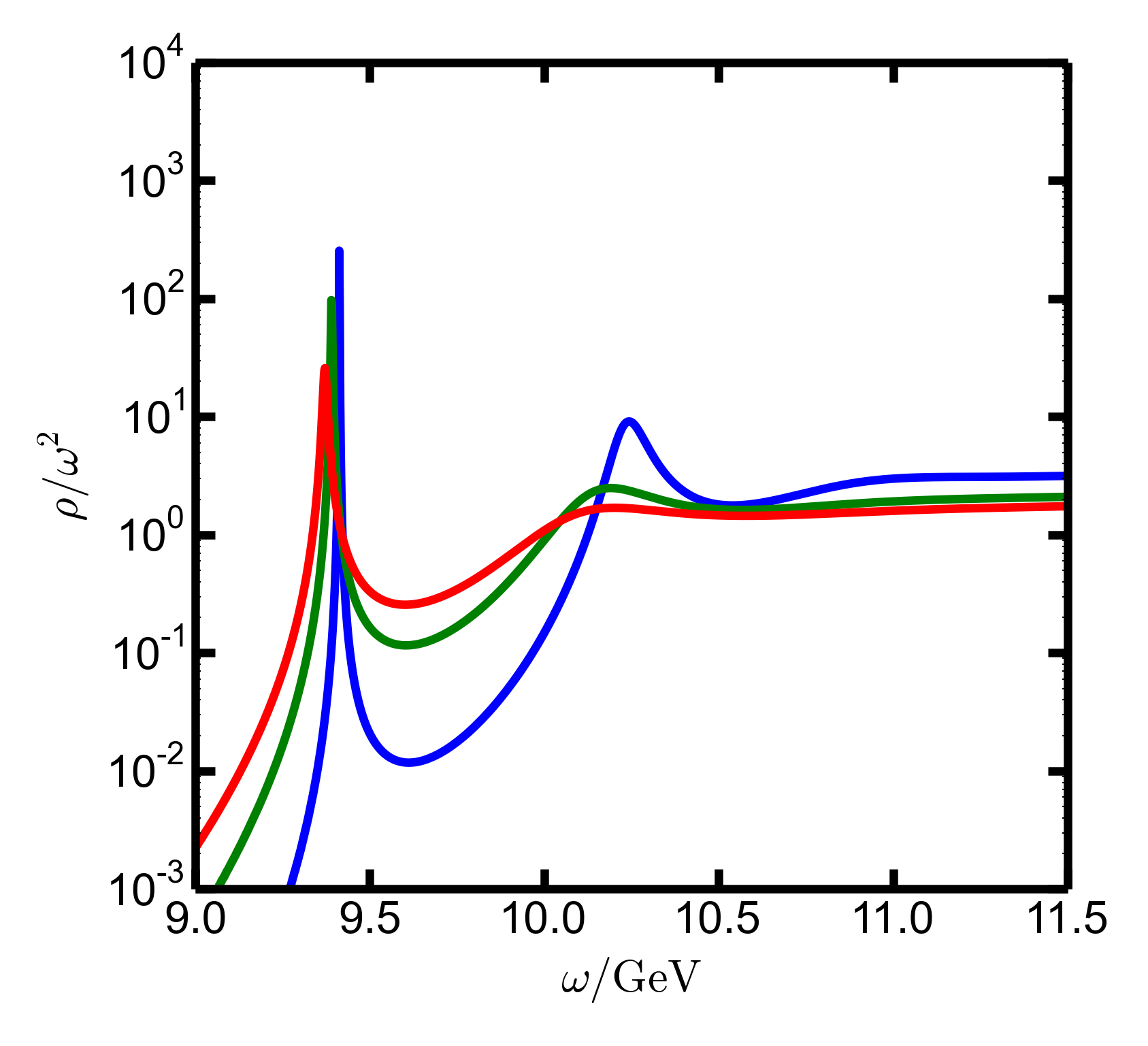}
\put(-383,107){$T = \, \, 145$~MeV} 
\put(-240,107){$T = 150$~MeV} 
\put(-90,107){$T = 155$~MeV}
\put(-79,50){{\color{blue} ----- $\mu_B = 0$}}
\put(-79,40){{\color{green} ----- $\mu_B = 100$~~MeV}}
\put(-79,30){{\color{red} ----- $\mu_B = 200$~~MeV}}
\caption{Scaled $\Upsilon$ spectral function $\rho (\omega, T, \mu_B) / \omega^2$
as a function of energy
$\omega$ for $T = 145$~MeV (left, i.e.\ at  $T/T_{v_s^2} = 1$), 
150~MeV (middle) and 155~MeV (right, i.e.\ at $T \approx T_{pc}$)
at $\mu_B = 0$ (blue), 100~MeV (green)
and 200~MeV (red). The data of Fig.~\ref{SF_Ups_crosssection_T} 
is rearranged to highlight the $\mu_B$ dependence and to furnish a comparison
with experimental results which use 
$M_{\mu^+ \mu^-}$ as abscissa.
\label{SF_Ups_crosssection_muB} 
}
\end{figure}

Our results are exhibited in Fig.~1 for the $\Upsilon$ meson. 
We emphasize that neither an explicit quark-mass dependence enters
our approach (instead, quark masses are implicitly accounted via 
$U_0(z)$ for entering $G_m$) nor a confinement criterion 
(instead, narrow spectral functions as quasi-particle states are
considered as confined $J^{PC} = 1^{--}$ $\bar b b$ states). In so far, 
the emergence of such narrow quasi-particle
ground states at $T = {\cal O}(T_{pc})$ is astonishing.
The higher the excitation, the later the excited-state formation happens 
when considering the cooling due to expansion. 
The net effect of the finite baryon-chemical potential $\mu_B$
is a lowering of the formation temperature.

The density of low-mass color carriers is $\propto T^3 + [\cdot] T^2 \mu_B$ in leading order of small $\mu_B$
and species-dependent positive constant $[\cdot]$,
i.e.\ increases with increasing $\mu_B$. In the spirit of the Matsui-Satz conjecture \cite{Matsui:1986dk},
the color charge screening
becomes stronger due to $\mu_B > 0$, and the formation of bound states is delayed (suppressed) 
therefore in a cooling medium.   
A closer look on the contour curves at $T = 140 \cdots 160$~MeV suggests
$\rho(\omega, T - 15~\mbox{MeV}, \mu_B = 0) \approx \rho(\omega, T, \mu_B = 200~\mbox{MeV})$,
meaning that the $\Upsilon$ formation pattern is shifted down by a temperature of about 15~MeV
by the impact of the baryo-chemical potential $\mu_B = 200$~MeV.
Otherwise, the minimum sound velocity, $T_{min \{ v_s^2\} } (\mu_B)$ drops only by about 5~MeV
when going from $\mu_B = 0$ to  $\mu_B = 200$~MeV. While being rather semi-quantitative and restricted to
$\Upsilon$,
this finding may be interpreted as a hint to $T_{fo} (\mu_b) \ne T_{pc} (\mu_b)$ at $\mu_B > 0$.

Focusing on the crucial temperature region near $T_{pc}$ or $T_{v_s^2}$, one observes how rapidly the ground state
evolves towards a sharp quasi-particle within this narrow interval of $T$ at $\mu_B = 0$, see left sharp peak in
left panel in Fig.~\ref{SF_Ups_crosssection_T}. The first excitation (the middle peak) becomes clearly visible,
with peak position noticeably shifting up upon dropping temperature. In contrast, the second excitation is identifiable
at $T = 145$~MeV but not so clearly at higher temperatures. These trends continue at $\mu_B > 0$, see middle and
right panels in Fig.~\ref{SF_Ups_crosssection_T}. At $\mu_B = 200$~MeV, the second excitation is not identifiable
as clear peak down to $T = 145$~MeV, while the first excitation sticks out only at $T \le 150$~MeV.
Let us emphasize that, at $\mu_B = 0$, the first and (weakly) the second excitations are identifiable as peak structures,
in contrast to \cite{MartinContreras:2021bis}, where these excitations appear as molten, while the ground state
persists up to high temperatures since it is kept by a narrow deep well potential.

To highlight the $\mu_B$ dependence, we exhibit in Fig.~\ref{SF_Ups_crosssection_muB} the same spectral functions
arranged in reversed order, i.e.\ various values of $\mu_B$ at a given temperature. Such a representation evidences the
impact of the baryo-chemical potential in a clear manner. Note that in an adiabatically cooling 
strong-interaction system one should employ the isentropic curves $T(\mu_B)$ to follow the evolution of the
spectral function. Figure \ref{SF_Ups_crosssection_muB} provides some guidance for that. 

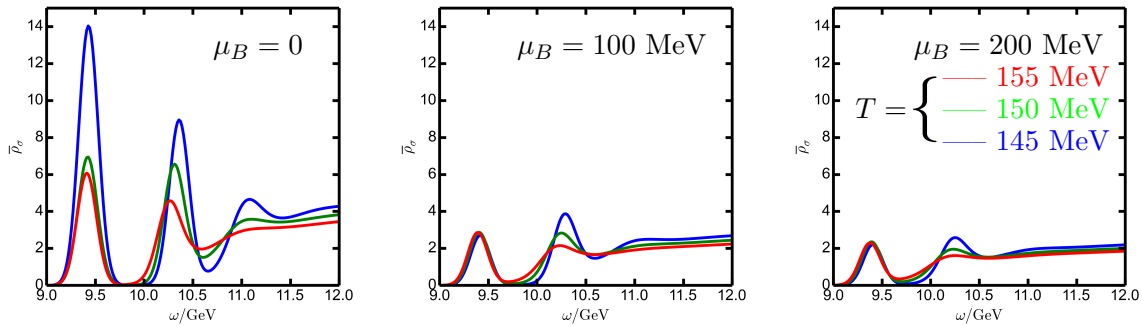
\begin{figure}[tb!]
\includegraphics[width=0.31\columnwidth]{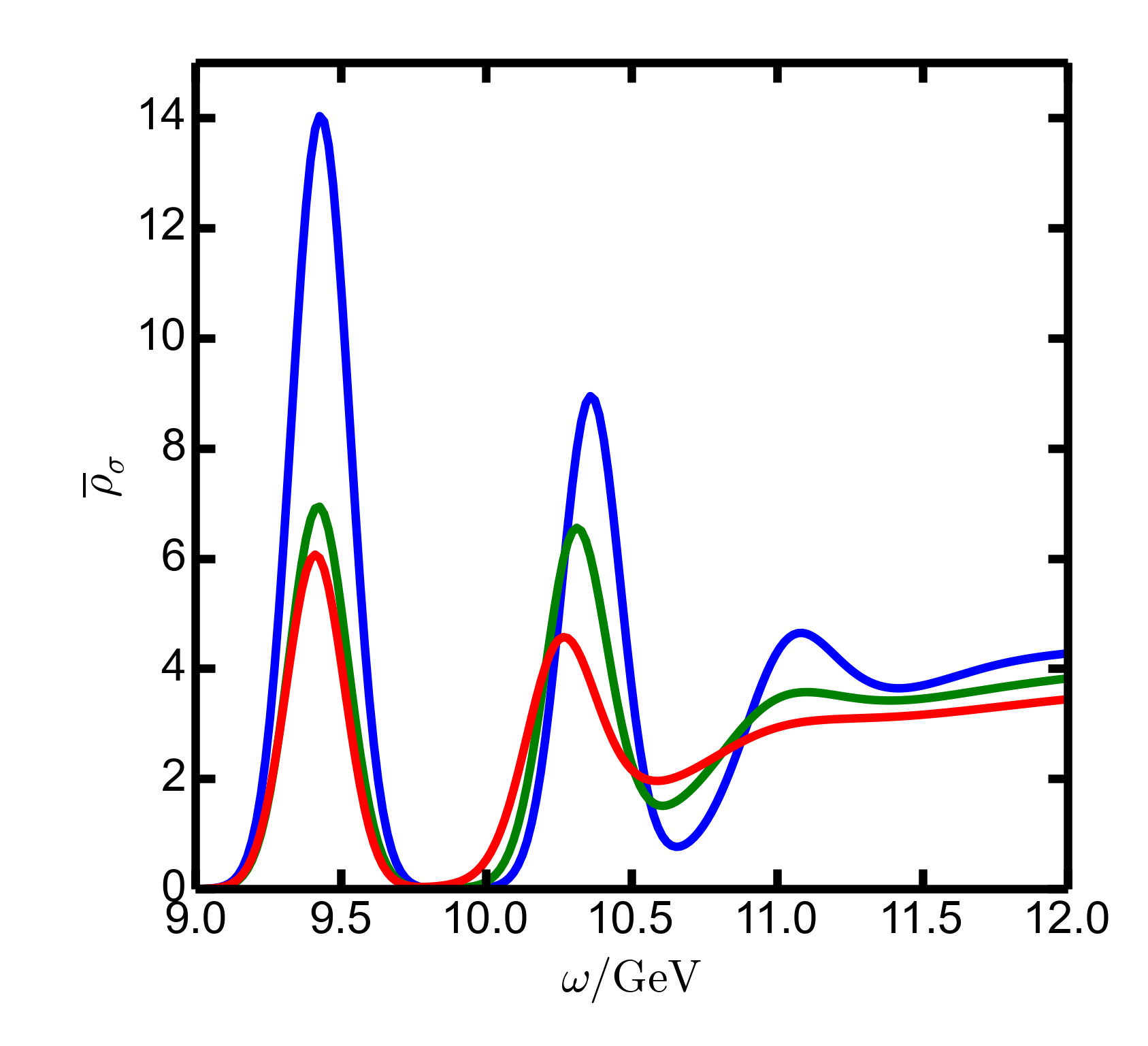}
\includegraphics[width=0.31\columnwidth]{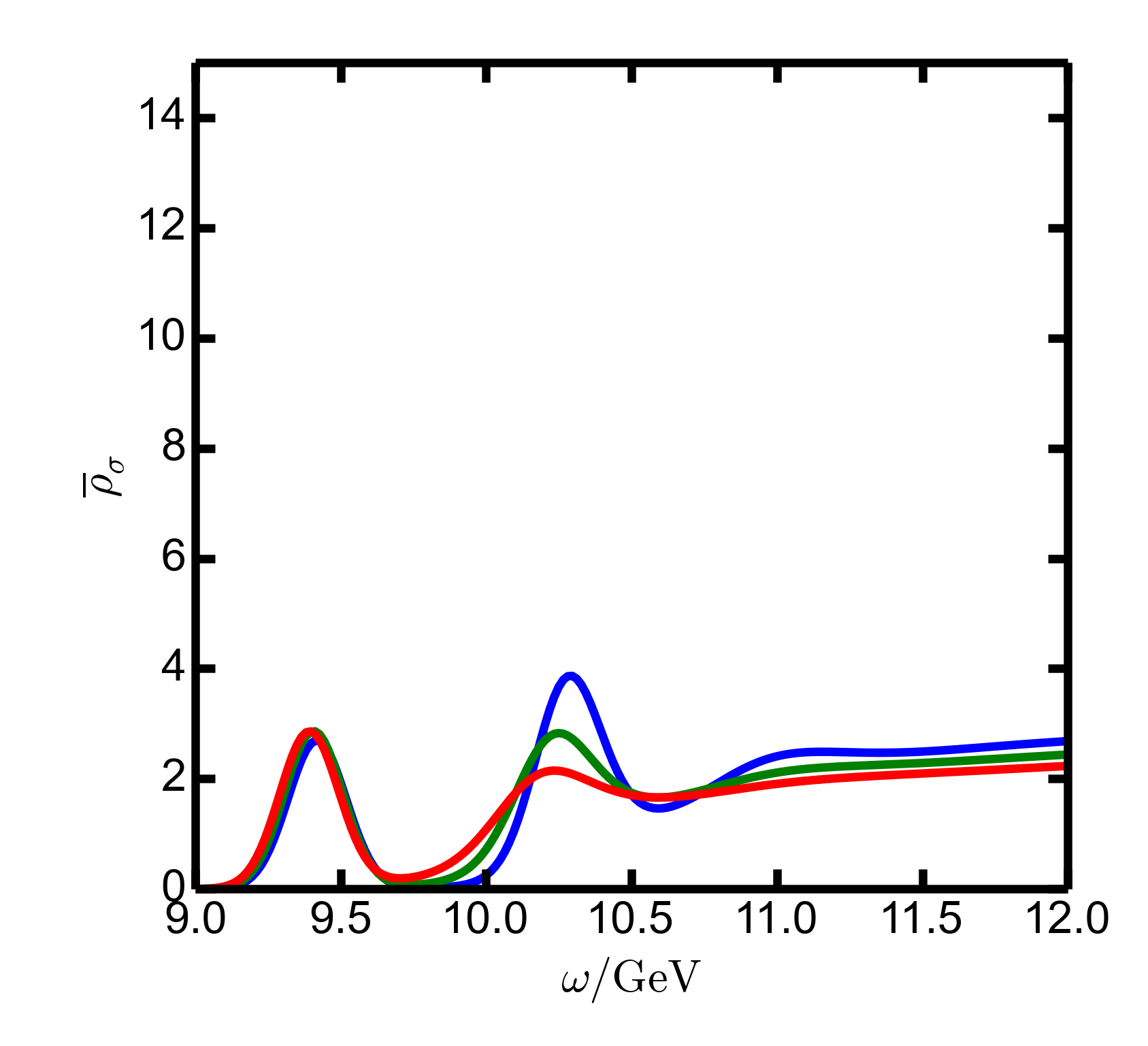}
\includegraphics[width=0.31\columnwidth]{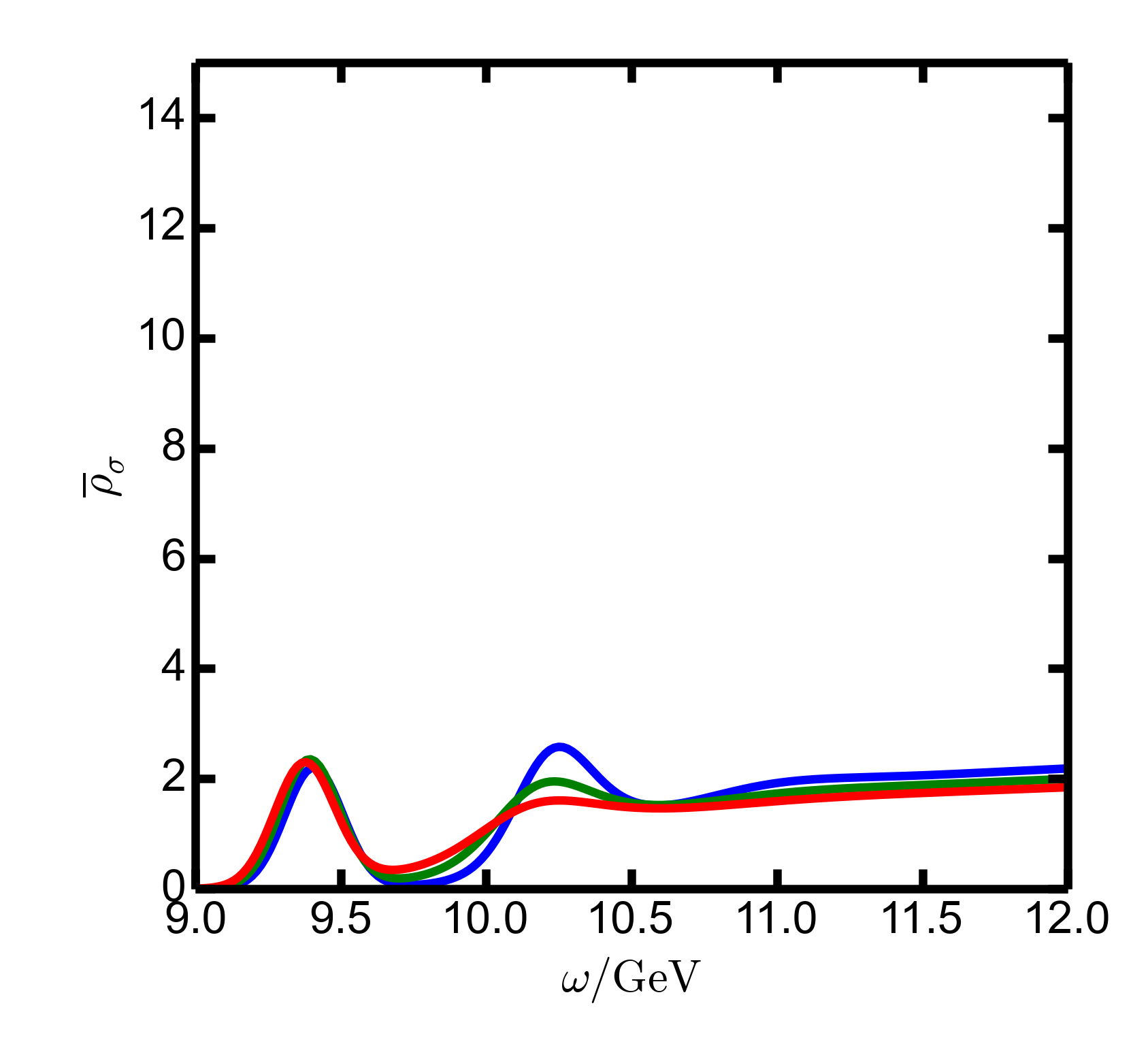}
\put(-355,107){$\mu_B = 0$} 
\put(-240,107){$\mu_B = 100$~MeV} 
\put(-90,107){$\mu_B = 200$~MeV}
\put(-79,95){{\color{red} ----- 155 MeV}}
\put(-79,83){{\color{green} ----- 150 MeV}}
\put(-79,71){{\color{blue} ----- 145 MeV}}
\put(-112,83){$T =$}
\put(-92,80){\Huge{$\{$}}
\caption{The same data as in Fig.~\ref{SF_Ups_crosssection_T} is used
but convoluted with a resolution function with width of 100~MeV (see the text)
to enable a comparison
with experimental results which use linear scales.
An adaptive $\bar \omega$ grid with minimum spacing of $10^{-8} L$ is employed.
Note that feeding is not included.
\label{SF_Ups_crosssection_muB_Gauss}  }
\end{figure}

The relation of the spectral function to the resulting $\mu^+ \mu^-$ spectrum from $\Upsilon \to  \mu^+ \mu^-$
may be elaborated as in previous studies, e.g.\ by superimposing the thermal yield (which needs a model of the
space-time evolution of the fireball) and the post-freeze-out contribution (which is directly related to the 
$\Upsilon(nS)$ yields and feedings) and the various background sources. This is beyond the scope of our paper.
Nevertheless, the emerging picture of our model (see $\mu_B = 0$ curves in Fig.~\ref{SF_Ups_crosssection_muB})
appears at the fist glance qualitatively consistent 
with experimental observations \cite{Chatrchyan:2012lxa,Sirunyan:2017lzi,Sirunyan:2018nsz,Acharya:2018mni}.
The strengths of excited states, $\Upsilon (2S, 3S)$, 
are gradually suppressed w.r.t.\ to the ground state  $\Upsilon (1S)$
in heavy-ion collisions with participant numbers $N_{part} > 100$, most notably the $3S$ state, 
while in $p p$ collisions one clearly identifies $\Upsilon (2S, 3S)$ as prominent peaks,
albeit with decreasing strengths.
One could imagine that convoluting our spectral functions with a finite (fiducial) resolution 
leads to a picture resembling better the observations,
e.g.\ figure 1 in \cite{Chatrchyan:2012lxa}.
In fact, applying a Gaussian resolution function according to the scheme
$\bar \rho_\sigma (\omega) = \int_0^\infty \dd \bar \omega \, \bar \rho (\bar \omega) \frac{1}{\sqrt{2 \pi \sigma^2}}
\exp \{ - \frac{(\bar \omega - \omega)^2}{ 2 \sigma^2} \}$ with $\bar \rho (\omega) := \rho(\omega)/\omega^2$
and selecting the {\it ad hoc} value 
$\sigma = 0.1$~GeV 
generates a pattern closer to the observation,
see Fig.~\ref{SF_Ups_crosssection_muB_Gauss}.
While the underlying soft-wall holographic potential captures approximately the mass spectrum of radial
excitations at $T = 0$, it facilitates, also in the present background,
decay constants increasing with (or independent of) radial quantum number $n$, 
in contrast to experimental data.
This imperfection seems to continue to $T > 0$: the strengths of
(i.e.\ yields from) excitations become too large. Therefore,  
we do not introduce a continuum or background  subtraction, as discussed in 
\cite{MartinContreras:2021bis}, and leave further refinements 
(e.g.\ the options offered in Appendix \ref{App:B3} w.r.t.\ decay constants)
and feeding corrections to follow-up work. 

\section{Conclusion and Summary}\label{summary}

Using a bottom-up holographic model with minimalistic field content, 
we investigate the
impact of the finite baryo-chemical potential $\mu_B$ on bottomonium formation
at temperatures in the order of the hadron-chemical freeze-out in relativistic
heavy-ion collisions.
The model has two pillars: (i) the vector meson part which employs the
bottomonium masses of ground state and radial excitations as 
input to adjust a suitable Schr\"odinger-equivalent potential, and (ii) the
Einstein-dilaton-Maxwell background which is adjusted independently to
lattice-QCD thermodynamics (sound velocity and light-quark susceptibilities).
The field content is as follows: 
(i) a bottomonium-specific function $G_m(\phi)$,
which encodes implicitly the $b$ quark masses via the Schr\"odinger-equivalent
potential $U_0(z)$ and is essential for the bulk-to-boundary propagator $\varphi (z)$,
and 
(ii) the gravity-dilaton-Maxwell part, determined dynamically by the dilaton potential
$V(\phi)$ and the dilaton-Maxwell coupling ${\cal F}(\phi)$. 
Since there is neither a confinement criterion nor a chiral condensate as order parameter
in such an approach,
we consider the shrinking of the spectral function $\rho$ (determined by  $\varphi$)
to a narrow quasi-particle state
as bottomonium formation in a cooling strong-interaction medium.
Despite a simple two-parameter Schr\"odinger-equivalent potential $U_0$, 
we find the bottomonium ground-state formation at about a temperature
of 150~MeV at $\mu_B = 0$.
Increasing $\mu_B$ drops the formation temperature. 
Excited states are consecutively formed at lower temperatures, or the
spectral strengths are not yet concentrated completely at the $T = 0$ quasi-particle energy at a given temperature.
This fits well in the experimental observation that the $\Upsilon (2S)$ and $\Upsilon(3S)$ states are hardly
identifiable in the di-lepton spectra in heavy-ion collisions ate LHC, 
while the ground state is clearly visible \cite{Chatrchyan:2012lxa}.
In contrast, $\Upsilon (1S, 2S, 3S)$ are clearly seen in proton-proton collisions
at the same beam beam energies per nucleon. 

Our approach assumes rapid thermalization and equilibration, 
since the cooling of the medium is handled as sequence of equilibrium states. 
Off-equilibrium phenomena up to dynamical freeze-out need to be considered
in refined investigations.
Highly desirable would be a closer contact to string theory
to overcome the deployed phenomenological parameterizations
steering our two pillars, background thermodynamics and vacuum mass spectrum.

\begin{appendix}

\section{Using the bulk coordinate $\mathbf{z}$ in the EdM model} \label{z_coordinate}

The equations of motion for the dilaton $\phi(z)$,
the Maxwell field $\Phi (z)$ with the coupling function 
${\cal F} (\phi)$, following from the action (\ref{EdM_action}),
read in the coordinates (\ref{eq:3}) 
with warp factor $\exp\{ A(z)\}$ and blackening function $f(z)$
\begin{eqnarray} \label{A1}
\phi'' &=& -[\frac32 A' + (\log f)'] \phi' 
- \frac{1}{f} [e^A \frac{\partial V(\phi)}{\partial \phi}
+ \frac18 \frac{\partial {\cal F}(\phi)}{\partial \phi} \Phi'^2], \\
\Phi'' &=& - [\frac12 A' + \frac14\frac{\log {\cal F}(\phi)}{\partial \phi} \phi' ] \Phi', \\
A'' &=& \frac12 A'^2 - \frac13 \phi'^2, \\
f'' &=& -\frac32 A' f' - \frac14{\cal F}(\phi) e^{-A} \Phi'^2 . \label{A4}
\end{eqnarray}
The leading-order initial conditions are (i) near boundary, i.e.\ $z \to 0^+$,
\begin{eqnarray} \label{A5}
\lim_{z \to 0 } \phi &=& 0, \hspace*{29mm} \lim_{z \to 0 } \phi' = 0, \\
\lim_{z \to 0 } \Phi &=& \mu_B L  , \label{A88} \\
\lim_{z \to 0 } A &=& - 2 \log \frac{z}{L} + \cdots, \quad  \lim_{z \to 0 } A' = - \frac{2}{z} + \cdots,\\
\lim_{z \to 0 } f &=& 1,  \label{A8}
\end{eqnarray}
and (ii) near horizon, i.e.\ $z \to z_H^-$, 
\begin{eqnarray}
\lim_{z \to z_H} \Phi &= 0 ,   \label{A10} \\
\lim_{z \to z_H} f &= 0 . \label{A9}
\end{eqnarray}
The Eqs.~(\ref{A10}) and (\ref{A9}) make these equations a 
mixed boundary problem.
The Hawking temperature is determined by 
$T(z_H) =  - \frac{1}{4 \pi} f(z, z_H)'\vert_{z=z_H}$ with a 
freely chosen value
of the horizon position $z_H$ and free choice of $\mu_B$.
For completeness,
we note also entropy density 
$s (T, \mu_B) = \frac{2 \pi}{\kappa_5^2} \exp\{ \frac32 A(z_H) \}$ 
and baryon density $n_B (T, \mu_B) = - \frac{1}{\kappa_5^2} \Phi_2 $
with $\Phi_2$ from the small-$z$ expansion $\Phi = \mu_B  L + \Phi_2 (z / L)^2 + \cdots$.

References \cite{Knaute:2017opk,Knaute:2017lll,Critelli:2017oub,Grefa:2021qvt}
use essentially the coordinates originally employed in 
\cite{DeWolfe:2010he,DeWolfe:2011ts} with special gauging of the radial coordinate.
These solutions can be parameterized by the double 
$(\phi(z_H), \Phi(z_H))$ and need {\it a posteriori} the determination
of the (screwed) $T$-$\mu_B$ mesh.
The advantage of Eqs.~(\ref{A1} - \ref{A9}) is in the boundary conditions
(\ref{A8}) and (\ref{A10}), which make easier the scan of the $T$-$\mu_B$ plane. 

\section{UV-IR matching} \label{UV_IR}

\subsection{Generalities}

In the zero-temperature limit, $T \to 0$ at $\mu_B = 0$, one has 
$f \to 1$, $\xi \to z$ and $U \to U_0$  in Eq.~(\ref{eq:6}).
The primary request to any useful Schr\"odinger-equivalent potential
$U_0(z)$ is to get the proper level spacing via
\begin{equation} \label{S_eq.}
y'' - (U_0 - \hat E L^{-2}) y = 0.
\end{equation}
A constant common shift of $U_0 \to U_0^b = U_0 + 4 b/L^2$ 
can be absorbed in $\hat E \to \hat E^b = \hat E + 4 b$
to accomplish the wanted meson ground-state mass squared, 
$L^2 m_0^{(b) \, 2} = \hat E_0  + 4 b$,
independent of $U_0(z)$. Here, we suppose that Eq.~(\ref{S_eq.}) delivers
a set of discrete eigenvalues $\hat E_n \equiv L^2 m_n^2$, $ n = 0, 1, 2, \cdots$ by the requirement of square-integrable
solutions $y_n$.

We emphasize again hat $U_0(z)$ is an independent input in our approach which determines ${\cal S}_0$,
via ${\cal S}_0'{}^2 + \frac12 {\cal S}_0^2 - 2 U_0 = 0$ from Eq.~(\ref{eq:7}), and $G_m(\phi_0)$, 
via $(\log G_m)' - {\cal S}_0 + \frac12 A_0' - \phi_0' = 0$ from Eq.~(\ref{eq:G}).
(The needed holographic background quantities $A_0(z)$ and $\phi_0(z)$ are determined independently
by the dilaton potential $V(\phi) = V(\phi_0)$, where the quantities at $T = 0$ and $\mu_B = 0$ are labeled by 
the subscript ``0".) That $G_m(\phi) = G_m(\phi_0)$ determines via Eqs.~(\ref{eq:EoM}) and (\ref{Green_function})
the spectral function. In so far, the choice of $U_0$ deserves some special attention.
 
\subsection{Approximately uncovering the \boldmath{$\Upsilon$} mass spectrum}\label{Ups_approx}

The famous soft-wall (SW) model \cite{Karch:2006pv} employs
$U_0^{SW}  = U_0^{UV} + U_0^{IR} = \frac34 z^{-2} + (a / L)^2 (z /L)^2$
with the leading-order asymptotic parts
\begin{eqnarray} \label{U_0_UV}
\lim_{z \to 0} U_0 \to U_0^{UV}(z) &:=& \frac{\alpha^2}{z^2} , 
\quad \alpha^2 \equiv \frac34 , \\
\lim_{z \to \infty} U_0 \to U_0^{IR}(z) &:=& \frac{a^2}{L^2} \left( \frac{z}{L} \right)^2 .
\label{U_0_IR}
\end{eqnarray}
It has one free parameter, $a$, and, in general,  
cannot accommodate independently ground state
mass and level spacing at the same time. Nevertheless, 
it delivers via $L^2 m_n^2 = 4 a (n + 1)$,
$n = 0, 1, 2 \cdots$, the Regge type
mass spectrum -- in  \cite{Karch:2006pv} termed ``linear confinement". 
Despite the imperfection, it has been
used in \cite{Fujita:2009wc,Fujita:2009ca} for an investigation of the termal behavior of the $J/\psi$ spectral function.
Supplemented with the shift parameter $b$, 
i.e.\ $U_0^{SW} \to U_0^{SW, b} = U_0^{SW} + 4 b/L^2$,
however, the ground state mass and uniform level spacing can be tuned separately and may be used as minimum parameter
model with $\hat E_n^{(b)} \equiv L^2 m_n^{(b) \, 2} = 4 b + 4 a (n +1)$.
The decay constants are less perfectly reproduced, as stressed in  
\cite{Grigoryan:2010pj} for $J/\psi$ and $\psi'$.
Nevertheless,
due to its transparency we stay with this variant in our study of
$\Upsilon$ spectral function. The parameters
$a = 0.2006$ and $b = 0.6436$ together with the scale setting 
$L^{-1} = 5.148$~GeV, used in Section \ref{num_results}, result in
$m_0^{(b)} = m_{\Upsilon (1S)} = 9.460$~GeV,
$m_1^{(b)}  = 10.524$~GeV ($= m_{\Upsilon(2S)} +5$\%) 
and $m_2^{(b)}  = 11.490$~GeV
($= m_{\Upsilon(3S)} + 11$\%).
A readjustment $a \to 0.1035$ and $b \to 0.7407$ would lead also to
the exact experimental mass of $m_1^{(b)}  = m_{\Upsilon (2S)} = 10.0233$~GeV 
as well as 
$m_2^{(b)}  = 10.556 $~GeV ($= m_{\Upsilon(3S)} + 2$\%) 
to be compared to $m_{\Upsilon (3S)} = 10.3553$~GeV,
but reduce somewhat the formation temperature, 
as discussed in \cite{Zollner:2020nnt}.
The below discussed dependencies on additional parameters
can be used for fine-tuning
by breaking the uniform level spacing $\hat E_{n+1} - \hat E_n = 4 a$ 
of the soft-wall model.

The relation $a \ll b$ suggests a separation of scales. The level spacing,
$m^2(\Upsilon((n+1)S)) - m^2(\Upsilon((nS)) \ll \frac12 ( m^2(\Upsilon((n+1)S)) + m^2(\Upsilon((nS)))$
for $n = 0, 1, 2$ 
may be attributed to QCD dynamics, while the  mass gap or average hadron masses squared
may be related to the heavy-quark mass. This calls for a separate consideration of the level spacing
as part of fine-tuning in Subsection \ref{App:B3}.

In attempting fine-tuning, one may proceed in a two-step approach by (i) first accomplishing the level spacing only,
and (ii) eventually shift the whole spectrum
to accomplish the wanted values of $m_n^{(b)}$. Applied to  $U_0^{SW, b}$, 
step (i) would fix $a$, and $b$ is obtained
in step (ii). While such a two-step fine-tuning procedure looks promising, it could be hampered by a problem which we
faced, e.g.\  in \cite{Zollner:2020cxb,Zollner:2020nnt}: unfavorable parameterizations of $U_0(z)$ can lead to too
low formation temperatures, such that at $T_{pc}$ quasi-particles are not yet formed,
in contrast to the common understanding of hadron formation in relativistic heavy-ion collisions
discussed in the introduction.
The origin of the affair can be qualitatively explained within the transparent model $U_0^{SW, b}$.
At finite temperatures, $U_0(z) \to U_T(z, z_H (T))$. Since $U_T (z, z_H) \propto f(z, z_H)$ according to 
Eq.~(\ref{eq:7}), one can imagine $U_T(z, z_H) \approx U_0(z) \Theta (z_H - z)$.
To accommodate the ground state in such a potential, the IR turning point (t.p., in the spirit of WKB)
$z^{IR \, t.p.} \approx 2 \sqrt{a + b} / a$  must obey
$z^{IR \, t.p.} <  z_H \approx 1/(4 \pi T)$. In other words, to allow for an ``unmolested" state at given
temperature $T$, the parameter $a$ must be sufficiently large to get small $z^{IR \, t.p.}$. 
This is the reasoning of considering the quark-mass effect encoded in $m_0$ as primary quantity
and a less strict parameter adjustment for the level spacing, as deployed in the above parameter setting. 
(In stark contrast, \cite{Braga:2018hjt} puts emphasis on the correct 
decay constants and is less restrictive to the bottomonium mass spectrum
with the advantage of rather persistent states up to high temperatures $T > T_{pc}$.)

\subsection{Fine-tuning of \boldmath{$U_0$} to recover $\Upsilon(1S, 2S, 3S)$ masses
by proper level spacing}\label{App:B3}

The asymptotic parts at small $z$ (UV), Eq.~(\ref{U_0_UV}), and large $z$ (IR), Eq.~(\ref{U_0_IR}), 
can be joint in many different ways
to a common Schr\"odinger-equivalent potential $U_0(z)$ 
to be used in Eqs.~(\ref{eq:6}) and, in vacuum, (\ref{S_eq.})
to accomplish the wanted fine-tuning.
Here we mention only one with a minimum set of parameters.
An easy choice would be
\begin{equation} \label{disc_pot}
U_0^{dip} =  \left\{
\begin{array}{l}
U_0^{UV}(z) + b^{UV} L^{-2} \, \mbox{ for } z < z_0,\\
\tilde U_0 \, \mbox{ for } z_0 \le z \le \lambda z_0 , \\
U_0^{IR}(z) + b^{IR}  L^{-2} \, \mbox{ for } z > \lambda z_0,
\end{array}
\right.
\end{equation}
with constant parameters $b^{UV, IR}$, $\tilde U_0$ , $\lambda$
and scale setting parameter $L$.
The options $\lambda = 1$ (dip related to discontinuity at $x_0$),
$\lambda \to 1$ (mimicking a Dirac delta dip at $x_0$ when $\tilde U_0 \propto 1/(\lambda -1 )$) and 
$\lambda > 1$ (box-like dip within $x_0 \cdots \lambda x_0$)
w.r.t.\ fine-tuning of the mass spectrum
are discussed in the Supplemental Material. 
We finish this essay by the expectation that the tendency
of the $\mu_B$ dependence of spectral functions is not obstructed by the details
of approximately or accurately adjusting parameters of $U_0$
to the $\Upsilon$ mass spectrum.  
 
\end{appendix}

\begin{acknowledgements}

The authors gratefully acknowledge the collaboration with J.~Knaute
and thank
M.~Ammon, P.~Braun-Munzinger, R.~Critelli,
M.~Kaminski, J.~Noronha, K.~Redlich and G.~R\"opke for useful discussions.
The work is supported in part by the European Union’s Horizon 2020 research
and innovation program STRONG-2020 under grant agreement No 824093. 

\end{acknowledgements}


\section*{Supplemental Material: Discussion of the Schr\"odinger-equivalent potential (\ref{disc_pot})}

The investigation of the quarkonium spectrum in \cite{Ebert:2011jc}, based on the relativistic quark model
with quasi-potential for parameterizing the $\bar Q Q$ interaction by a Schr\"odinger type equation
(not to be messed up with our Schr\"odinger-equivalent equation (\ref{eq:6}) which arises by transforming
a second-order equation of motion of the vector field ${\cal A}$ into a convenient form) unravels a non-linear
$\Upsilon$ Regge trajectory of radial excitations deviating clearly from the linear regression fit
$m_{n \, fit}^2 = 6.536 \, \mbox{GeV}^2 (n -1) + 92.157 \, \mbox{GeV}^2$ 
with $m_{\Upsilon (1S)} < m_{0 \, fit}^2$ and  $m_{\Upsilon (2, 3S)} > m_{1,2 \, fit}^2$
(see figure 7 in \cite{Ebert:2011jc}). 
The way of recalling such facts supports our below handling of the
model parameters, which either approximately (Subsection \ref{Ups_approx})
or accurately (Subsection \ref{App:B3})
reproduce the experimental 
$\Upsilon (n S)$ $I^G (J^{PC}) = 0^- (1^{--})$ mass spectrum in vacuum, i.e.\ at
$T = 0$ and $\mu_B = 0$.

For $\lambda = 1$, Eq.~(\ref{disc_pot}) displays, 
in general, a discontinuity at $z_0$, which could be removed
by suitable values of $b^{UV, IR}$ dependent on $z_0$:
$z_0^2 = \frac{L^2}{2 a} (b^{UV}-b^{IR} \pm \sqrt{(b^{UV}-b^{IR})^2 + 3 a})$.  
In \cite{Grigoryan:2010pj}, such a discontinuity
is admitted, even amplified by an additional negative Dirac distribution at $z_0$
(the ``dip", which demolishes the smoothness of $y'$ at $z_0$, 
since $y'(z_0^+) - y'(z_0^-) = - y(z_0) \int_{z_0^-}^{z_0^+} \dd z \, U_0(z) \ne 0$
for $y \in C^0$ and $U_0$ piece-wise continuous;
the Dirac delta dip can be mimicked by Eq.~(\ref{disc_pot}) by the limiting procedure
$\lambda \to 1$ and $\tilde U_0 =  \Delta / (\lambda - 1)$
being equivalent to $\Delta \delta(z-z_0)$ for 
$z \in [z_0 - \epsilon, z_0 + \epsilon]$ at $\epsilon /z_0 \ll 1$.) 
The emerging four-parameter ``dip and shift" potential
in \cite{Grigoryan:2010pj} was adjusted to $m_{J/\psi, \psi'(2S)}$ together with decay constants,
see also \cite{Hohler:2013vca}. We focus on the $1S$, $2S$, and $3S$ 
bottomonium masses only and include in $U_0^{dip}$
the finite-width dip $\propto (\lambda - 1)$ of depth $\tilde U_0$.
For a reduction of parameters we put henceforth
$b^{UV, IR} = 0$. Even then, the potential (\ref{disc_pot})
displays for $\lambda = 1$, in general, a discontinuity at $z_0$ with
a dip at l.h.s.\ (r.h.s) of $z_0$ for $z_0 < (>) (\frac34)^{1/4} a^{-1/2}$.
The dip becomes pronounced for $\lambda > 1$; we suppose
$\tilde U_0 < min\{U_0^{UV} (z_0), U_0^{IR}(\lambda z_0) \}$.
Rather than investigating the results for a particular point in parameter space, 
we are interested in the systematic enabled by the ansatz (\ref{disc_pot}).

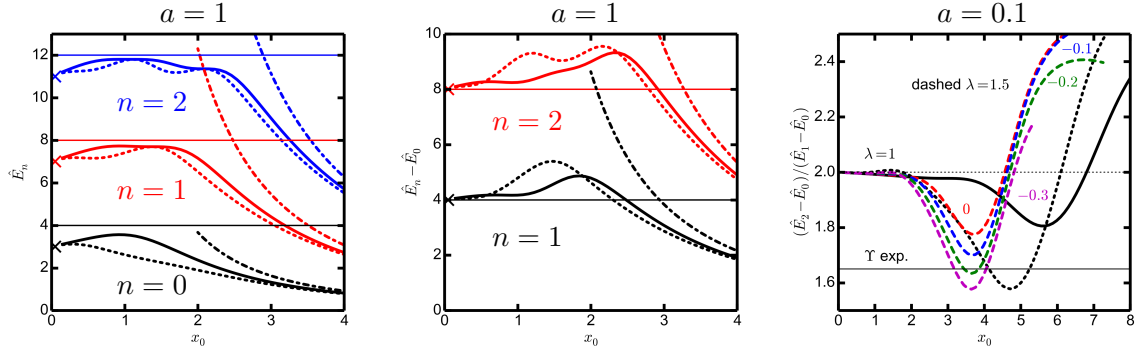
\begin{figure}[t!]
\includegraphics[width=0.31\columnwidth]{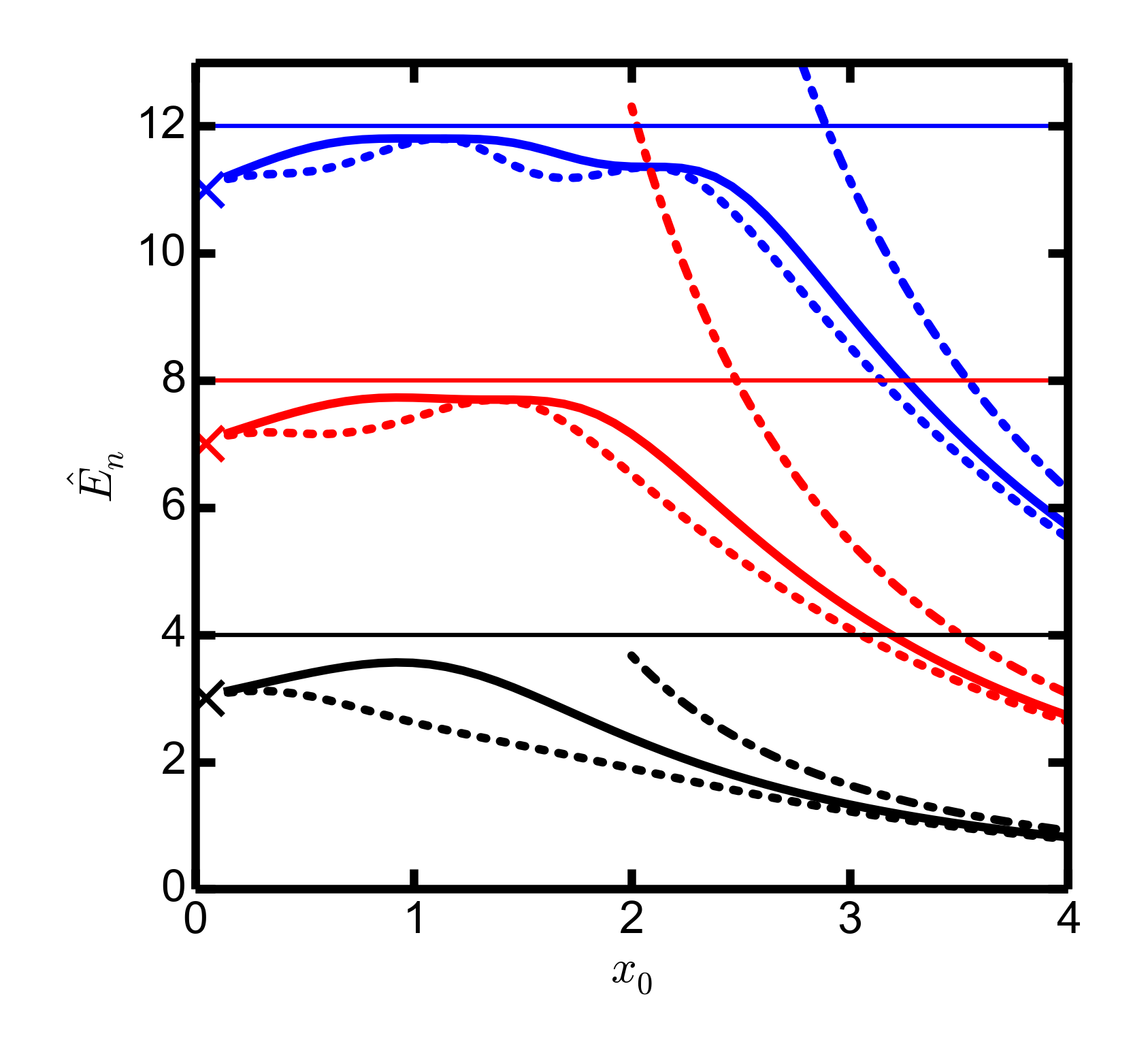} 
\includegraphics[width=0.31\columnwidth]{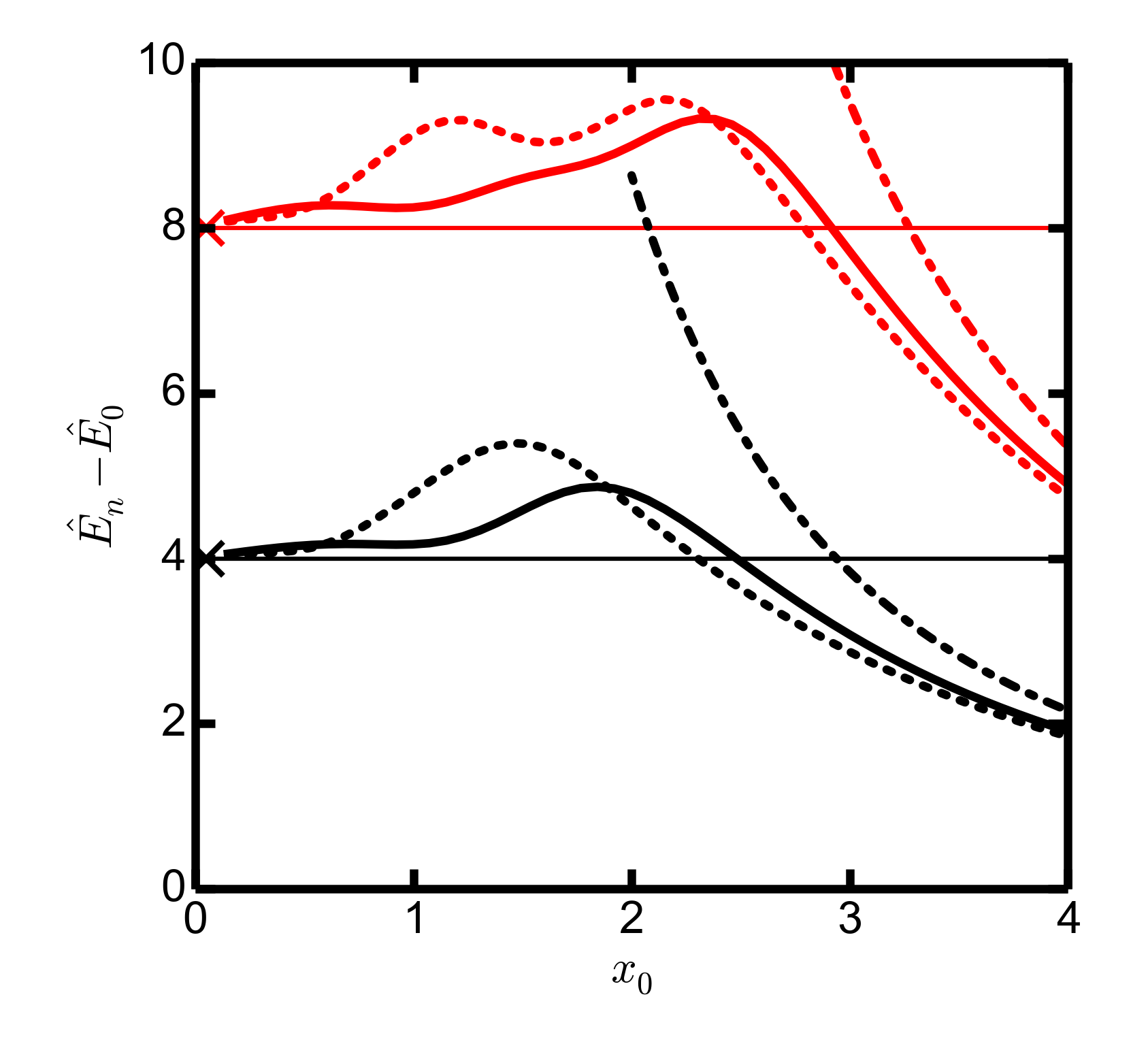} 
\includegraphics[width=0.31\columnwidth]{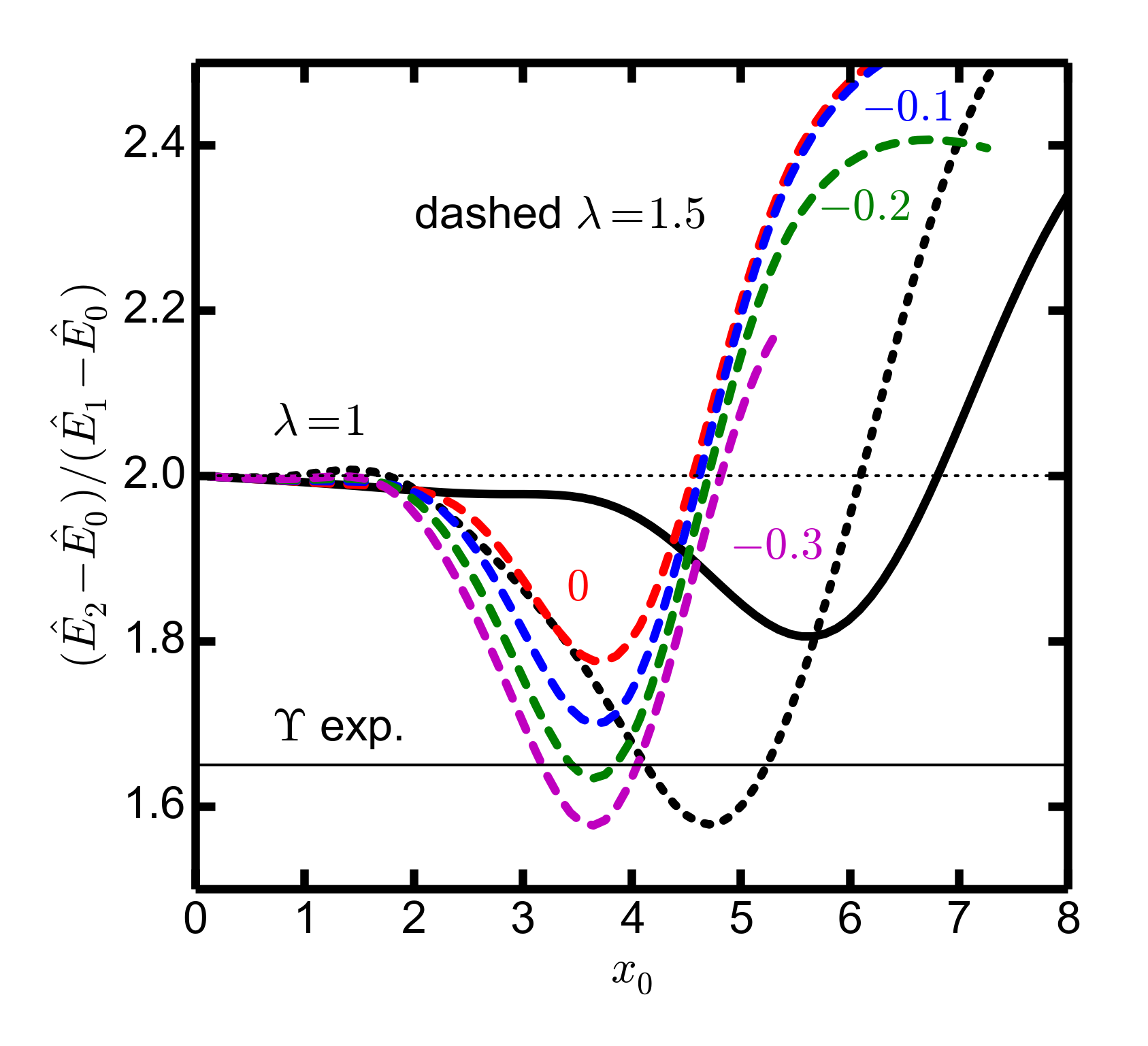} 
\put(-377,129){$a = 1$}
\put(-229,129){$a = 1$}
\put(-83,129){$a = 0.1$}
\put(-397,26){ {\color{black} $n=0$}}
\put(-397,62){ {\color{red} $n=1$}}
\put(-397,95){ {\color{blue} $n=2$}}
\put(-255,44){ {\color{black} $n=1$} }
\put(-255,88){ {\color{red} $n=2$} }
\caption{Survey on the properties of energy eigenvalues resulting from
the Schr\"odinger-equivalent potential (\ref{disc_pot}) 
under variation of the UV-IR matching point $x_0$ 
for the three options $\lambda = 1$ (discontinuity at $x_0$),
$\lambda \to 1$ (Dirac delta at $x_0$) and 
$\lambda > 1$ ( box-like dip within $x_0 \cdots \lambda x_0$). 
Left panel: The first energy eigenvalues 
$\hat E_n = L^2 m_n^2$ 
for $\lambda = 1$ (solid curves) and $\lambda \to 1$,
$\tilde U_0 = \Delta /(\lambda -1)$, $\Delta = - 0.5 L^{-2}$ (dotted curves). 
In addition, the crosses at $x_0 = 0$ are for the half-side harmonic oscillator, 
and the dashed curves at $x_0 > 2$
depict results of the r.h.s.\ hard wall at $x_0$. Thin horizontal lines are for the 
soft-wall model with $U_0^{SW} = U_0^{UV} + U_0^{IR}$. 
For $a = 1$.
Middle panel: As in left panel but for the level spacing, i.e.\
differences $\hat E_n - \hat E_0 = L^2 (m_n^2 - m_0^2)$. 
Right panel: Ratio $(\hat E_2 - \hat E_0)/(\hat E_1- \hat E_0) =
(m_2^2 - m_0^2)/(m_1^2 - m_0^2)$ 
for  $\lambda = 1$ (solid black curve) and 
$\lambda = 1.5$  (dashed curves), displayed only for $\hat E_n > 0$) 
with $L^2 \tilde U_0 = 0$ (red),
-0.1 (blue), -0.2 (green) and -0.3 (violet).
The black dotted curve is as in the left and middle panels for
the $\lambda \to 1$ model.
The thin black horizontal line displays 
$(m_{\Upsilon(3S)}^2 -  m_{\Upsilon(1S)}^2)/(m_{\Upsilon(2S)}^2 -  m_{\Upsilon(1S)}^2)$ with 
experimental masses, while the thin dotted horizontal line depicts the maximum ratio value
if one of the experimental masses is modified by $\pm 1$\%. For $a = 0.1$.
\label{UV_IR_matching}
}
\end{figure}

The solutions of the respective Schr\"odinger equations are as follows:
\renewcommand{\theequation}{\Roman{equation}}
\setcounter{equation}{0}
\begin{eqnarray}
y'' - (L^2 U_0^{UV} - \hat E) y &=& 0: \,
y = c_1^{UV} \sqrt{x} J_1(\sqrt{\hat E} x) + 
c_2^{UV} \sqrt{x} Y_1(\sqrt{\hat E} x), \label{y_UV} \\
y'' - (L^2 \tilde U_0 - \hat E) y &=& 0: \,
y = c_1^{dip} \exp\{ \sqrt{ L^2 \tilde U_0 - \hat E} x \} + 
c_2^{dip} \exp\{-\sqrt{ L^2 \tilde U_0 - \hat E} x \}, \label{y_dip}\\
y'' - (L^2 U_0^{IR} - \hat E) y &=& 0: \,
y = c_1^{IR} D_{\frac{\hat E - a}{2 a }} (\sqrt{2 a} x) + 
c_2^{IR} D_{-\frac{\hat E + a}{2 a }} (i \sqrt{2 a} x),  \label{y_IR} 
\end{eqnarray} 
where $J_1$ and $Y_1$ stand for the Bessel functions, and
$D_\nu$ are parabolic cylinder functions; 
$x \equiv z/L$ from here on.\footnote{\label{fussnote}
The Eqs.~(\ref{y_UV}, \ref{y_IR}) 
are limiting cases of the solution of 
$y'' -(\frac{\alpha^2}{x^2} + a^2 x^2 - \hat E) y = 0$,
\begin{equation} \tag{*}
y = x^{\frac{1 + W}{2}} e^{- \frac{a x^2}{2}} \, 
\left\{ c_1 U \left(\frac14 [2 +W - \frac{\hat E}{a}], \frac12 [2 + W], a x^2 \right)
+ c_2 L_{[ \frac{\hat E}{a} - 2 - W] / 4}^{W/2} \left( a x^2 \right) \right\} ,
\end{equation}
with (i) $W = \sqrt{1 + 4 \alpha^2}$
for $a = 0$ and $\alpha^2 = 3/4$, yielding (\ref{y_UV}), as well as (ii) 
$\alpha^2 = 0$, yielding (\ref{y_IR})  (see \cite{W_alpha} for the definition and properties
of the confluent hypergeometric (Kummer) function of second kind $U(a, b, x)$, also known as Tricomi function, and 
the associated Laguerre polynomial $L_\gamma^\rho (x)$).
Square integrability for $x \in (0, \infty)$ unravels the energy eigenvalues 
$\hat E = a( 4n + 2 \sqrt{\alpha^2+\frac14} +2)$ 
which become these of the soft-wall model for $\alpha^2 = 3/4$, 
i.e.\ $\hat E = 4 a (n+1)$.
Adding a Dirac delta at $x_\delta$ with strength $- \Delta$, the solutions (*) for $x< x_\delta$ and $x > x_\delta$
must be matched at $x_\delta$ with the above condition for the discontinuity of $y'$:
$y'(x_\delta^-) - y'(x_\delta^+) = y(x_\delta) \Delta$. 
The continuity condition of $y$ at $x_{\delta}$ determines one of the four integration constants of 
the general solution (*) with its two branches and the discontinuity condition of $y'$ fixes another. 
The remaining two constants are determined by the claim of square integrability and 
the related appropriate boundary conditions and the normalization of the solution in case of 
normalizability for all energy eigenvalues $\hat E$. 
In contrast to highly symmetric square-well/harmonic oscillator + 
Dirac delta models of
\cite{Belloni_Robinett,Dirac_delta}, 
the ground state energy cannot be dialed independently of the
excited states and their (non-uniform) level spacing. 
}
Given the asymptotic behavior (i)
$\lim_{x \to 0} Y_1(x) \to - 2/(\pi \, x) + \cdots$, 
the square-integrability requirement 
in $x \in (0, x_0)$ facilitates  $c_2^{UV} = 0$, and
(ii) since $D_{-\frac{\hat E + a}{2 a }} (i \sqrt{2 a} x) \propto \exp\{ a x^2 /2\}$ 
in leading order at large $x$,  $c_2^{IR} = 0$ ensures
square-integrability in $x \in (\lambda x_0, \infty)$.\footnote{
Since for $\nu \equiv \frac{\hat E - a}{2 a} = n$, $n = 0, 1 , 2 \cdots $,
the parabolic cylinder functions are related to Hermite polynomials $H_n$ via
$D_n(x) = 2^{-n/2} \exp\{ - x^2/4 \} H_n(x/\sqrt{2})$, and
Eq.~(\ref{y_IR}) becomes
\begin{equation}\tag{\ref{y_IR}'}
y = c_1^{IR} 2^{-n/2} \exp\{ - \frac{a}{2} x^2\} H_n (\sqrt{a} x) +
 c_2^{IR} 2^{-n/2} \exp\{ \frac{a}{2} x^2\} H_n (i \sqrt{a} x),
\end{equation}
again with $c_2^{IR} = 0$ upon square integrability. The boundary condition $y(x=0) =0$
yields the half-side harmonic oscillator.}
The solutions $y(\mbox{Eq.~(\ref{y_UV})})$, $y(\mbox{Eq.~(\ref{y_dip})})$ 
and $y(\mbox{Eq.~(\ref{y_IR})})$ and their logarithmic derivatives 
must be matched at $x_0 = z_0 /L$ and, if $\lambda > 1$, at $\lambda x_0$.
Properties of the eigenvalues $\hat E_n$
can be recognized in Fig.~\ref{UV_IR_matching}. 

Let us first consider the case of $\lambda = 1$, see left panel in Fig.~\ref{UV_IR_matching} which is for $a = 1$. 
The condition $x_0 \ll 1$, i.e.\ an altitude-limited l.h.s.\ hard wall,
facilitates corrections to the half-side harmonic oscillator,
$\hat E_n \approx a (4 n + 3 + \epsilon_n)$ with
$\epsilon_n = \frac{1}{6} \frac{(2 n + 2)!}{4^n n! (n+1)!} 
\sqrt{\frac{a}{\pi}} x_0$ of the small-$x_0$ expansion
by zeroes of $D_{\frac{\hat E - a}{2 a}} (\sqrt{2 a} x_0) = 0$ 
as a function of $\hat E$
for the truncated harmonic oscillator which is l.h.s.-limited by a hard wall at position $x_0 \ge 0$ .
The opposite case, $x_0 \gg 1$, is for an altitude-limited r.h.s.\ hard wall
yielding $\hat E_\ell \approx (x_\ell / x_0)^2$ with $x_\ell$ being the zeroes 
of $J_1$, $J_1(x_\ell) = 0$,
$\ell = 1, 2, \cdots$, see dashed curves. 
In both cases, the respective energies $\hat E_{n, \ell}$
must be smaller than the maxima of the altitude-limited hard walls:
$\hat E_n < L^2 U_0^{UV}(x_0^-)$ or $\hat E_\ell < L^2 U_0^{IR} (x_0^+)$.

Despite the noticeable change of the energy eigenvalues $\hat E_n$ with changing parameter $x_0$ 
(see solid and dotted curves in the left panel), 
the level spacing is less influenced, see solid and dotted curves in the middle panel.
The level spacing, given be differences of energy eigenvalues 
$\hat E_n - \hat E_0 = L^2 (m_n^2 - m_0^2)$ 
($\approx 4 a (n+1)$ for $x_0 < 1$, see middle panel), 
can be related to
the $\Upsilon ( 1S, 2S, 3S)$ mass-squared differences $m_n^2 - m_0^2$,
10.815~GeV${}^2$ ($n=1$) and 17.623~GeV${}^2$ ($n=2$),
pointing either to $a \approx 0.1$ ($n=1$) or $a \approx 0.08$ ($n=2$),
since $L^{-1} = 5.148$~GeV. Clearly, the choice $\lambda = 1$
only approximately accommodates the proper level spacing, as evidenced 
by two slightly different values of $a$. 

In fact, the ratio $\frac{\hat E_2 - \hat E_0}{\hat E_1- \hat E_0}$
(which is independent of $L^2$, $a$ and a shift by $b$)
quantifies better the level spacing. 
It can be directly related to the experimental
ratio $\frac{m_{\Upsilon(3S)}^2 -  m_{\Upsilon(1S)}^2}{m_{\Upsilon(2S)}^2 -  m_{\Upsilon(1S)}^2}$,
see the thin black horizontal line in the right panel. 
The variation of the matching point $x_0$ 
in the $\lambda=1$ model (\ref{disc_pot}) alone is not sufficient 
to meet exactly the experimental value, as pointed out above
and depicted by the dashed black curve in the right panel.
Tiny variations of one of the values of $m_{\Upsilon(nS)}$ 
on the 1\% level induce changes of the ratio (see dotted horizontal line) 
which are comparable with the change caused by $x_0$ variation.
The model  (\ref{disc_pot}) with $\lambda > 1$ and suitable value of 
$\tilde U_0$, however, is capable to accommodate any of the wanted
ratio value, see solid colored curves in the right panel for $a = 0.1$.
The choice
$\lambda = 1.5$ and various values of $L^2 \tilde U_0$ represent examples
of accomplishing the wanted fine-tuning. Analogously, the $\lambda \to 1$ model
with suitable value of $\Delta$ is also appropriate (see black dotted curve).

The case $\lambda \to 1$ means
replacing the $\tilde U_0$ dip in Eq.~(\ref{disc_pot}) for $x \in [x_0, \lambda x_0]$ by a Dirac delta,
$- \Delta \delta (x - x_0)$. 
As pointed out above and in footnote \ref{fussnote},
the solutions
$y(\mbox{Eq.~(\ref{y_UV})})$ and $y(\mbox{Eq.~(\ref{y_IR})})$ 
and their derivatives
must be matched at $x_0$ to yield 
\begin{equation} \label{UV_IR_match}
-\frac{1}{2 x_0} \left[1 - 2\sqrt{\hat E} x_0 \frac{J_0(\sqrt{\hat E} x_0)}{J_1(\sqrt{\hat E} x_0)} \right]
- \frac{\sqrt{2 a}}{2} \left[\sqrt{2 a } x_0 -
 2 \frac{D_{\frac{\hat E + a}{2 a}} (\sqrt{2 a} x_0)}{D_{\frac{\hat E - a}{2 a}} (\sqrt{2 a} x_0)} \right]
= \Delta
\end{equation}
for discrete energy eigenvalues $\hat E (x_0, a, \Delta)$ to be numerated by $n$, 
see dotted curves in Fig.~\ref{UV_IR_matching}.

\end{document}